# Quantum mechanical equivalence of the metrics of a centrally symmetric gravitational field

M.V.Gorbatenko[1], V.P.Neznamov [1,2*]

[1]FSUE RFNC-VNIIEF, Russia, Sarov, Mira pr., 37, 607188

[2]National Research Nuclear University MEPhI, Moscow, Russia

Abstract

We analyze the quantum mechanical equivalence of the metrics of a centrally symmetric uncharged gravitational field. We consider the static Schwarzschild metric in spherical and isotropic coordinates, the stationary Eddington-Finkelstein and Painlevé-Gullstrand metrics, and nonstationary Lemaître-Finkelstein and Kruskal-Szekeres metrics. When the real radial functions of the Dirac equation and of the second-order equation in the Schwarzschild field are used, the domain of wave functions is restricted to the range $r > r_0$, where $r_0$ is the radius of the event horizon. A corresponding constraint also exists in other coordinates for all considered metrics. For the considered metrics, the second-order equations admit the existence of degenerate stationary bound states of fermions with zero energy. As a result, we prove that physically meaningful results for a quantum mechanical description of a particle interaction with a gravitational filed are independent of the choice of a solution for the centrally symmetric static gravitational field used.

---
[*] vpneznamov@vniief.ru
vpneznamov@mail.ru

## 1. Introduction

The Schwarzschild metric [1] is a well-known solution of the general relativity (GR) theory. The Schwarzschild solution is characterized by a pointlike spherically symmetric source of the gravitational field of a mass $M$ and an "event horizon" (gravitational radius) $r_0 = 2GM/c^2$, where $G$ is the gravitational constant and $c$ is the speed of light. The surface of radius $r_0$ is usually regarded as a coordinate singularity. From the standpoint of a distant observer, a test particle reaches the event horizon in an infinite time. For the complete or partial elimination of the coordinate singularity of the Schwarzschild metric at the event horizon, the following solutions of GR were previously obtained: the Schwarzschild metric in isotropic coordinates ($S_{is}$) [2], the Eddington-Finkelstein (EF) metric [3], [4], the Painlevé-Gullstrand (PG) metric [5],[6], the Lemaître-Finkelstein (LF) metric [4], [7], and the Kruskal- Szekeres (KSz) metrics [8], [9]. These solutions can be obtained via the coordinate transformations of the static Schwarzschild metric. After these transformations, the Schwarzschild metric in isotropic coordinates remains static, the EF and PG metrics become stationary, the LF and KSz metrics become nonstationary.

In quantum mechanics, the motion of half-spin particles in an external nonquantized gravitational fields can be described by both complex and real radial wave functions of the Dirac equation. If we use complex functions, then the radial currents of the Dirac particles are nonzero, in particular, near the event horizon $r_0$. As a result, there is a "sink" of particles into the singularity at the coordinate origin $r = 0$. In this case, for static metrics (Schwarzschild and Schwarzschild in isotropic coordinates) and stationary EF and PG metrics, there are quasistationary (or resonance) states of the Dirac particles with complex energies decaying in time. Such states for massive scalar particles in the Schwarzschild field were considered in [10] - [13] using the Klein-Gordon equation. A similar problem for Dirac particles was studied in [14] - [18]. The problem of quasistationary states for half-spin particles in the PG field was considered in [19].

The situation changes qualitatively if we use real radial functions of the Dirac equation. In this case, the radial currents of Dirac particles are zero in the whole domain of wave functions [20]. From the realness condition for the equations for the real radial functions, the domain of wave functions for the Schwarzschild metric is $(r_0, \infty)$. But the radial functions are square nonintegrable as $r \to r_0$, and for nonzero values of the particle energy, the regime of the particle "fall" to the event horizon is realized [20] - [22]. The square integrability of wave functions near the event horizon can be restored if the fermion motion is described by second-order self-conjugate equations with spinor wave functions [20]. From the second-order equation, we can obtain a degenerate stationary bound state of a zero-energy massive fermion. A normalizable eigenfunction vanishing at the event horizon corresponds to this state. The effective potential of the second-order equation has a singularity at the event horizon, and this singularity admits stationary bound states of fermions.

Here, we answer the question of what happens with the degenerate eigenvalues $E = 0$ and with the fermion eigenfunctions under coordinate transformations of the Schwarzschild metric. The problems discussed here do not exhaust the wide subject area of spinning-particle dynamics in gravitational fields. We just recall the recent Gravity Probe B experiment with a gyroscope on board the satellite, which confirmed the GR predictions regarding two types of precession. At the same time some fundamental questions remain unclear. Such questions include spinning-particle behavior in the region of strong gravitational fields and the existence of the bound states "collapsar + fermion". We discuss these questions in this paper.

**2. Analysis methodology for the quantum mechanical equivalence of central symmetric solutions of GR equations**

**2.1. Dirac equation**

Half-spin particles motion in an external gravitational field is usually described in most papers by the covariant Dirac equation. The Dirac equation in the system of units with $\hbar = c = 1$ and in the signature $(+---)$ is

$$i\gamma^{\alpha}\nabla_{\alpha}\psi - m\psi = 0. \tag{1}$$

Here, $m$ is the particle mass, $\psi$ is a four-component bispinor, $\nabla_{\alpha}$ is the covariant derivative, and the $\gamma^{\alpha}$ are the Dirac world 4x4 matrices satisfying the relation

$$\gamma^{\alpha}\gamma^{\beta} + \gamma^{\beta}\gamma^{\alpha} = 2g^{\alpha\beta}E, \tag{2}$$

where $g^{\alpha\beta}$ is the inverse metric tensor and $E$ is the 4x4 unit matrix. Here and hereafter, quantities denoted by Greek letters take the values 0, 1, 2, and 3, quantities denoted by Latin letters take the values of 1, 2, and 3. Repeated upper and lower indices imply summation of the appropriate terms.

In what follows, along with Dirac matrices $\gamma^{\alpha}$ with world indices, we use the Dirac matrices $\gamma^{\underline{\alpha}}$ with local indices satisfying the relation

$$\gamma^{\underline{\alpha}}\gamma^{\underline{\beta}} + \gamma^{\underline{\beta}}\gamma^{\underline{\alpha}} = 2\eta^{\underline{\alpha}\underline{\beta}}E. \tag{3}$$

The quantity $\eta^{\underline{\alpha}\underline{\beta}}$ in (3) corresponds to the metric tensor of the Minkowski flat space with the signature $\eta_{\underline{\alpha}\underline{\beta}} = \text{diag}[1,-1,-1,-1]$. It is convenient to choose $\gamma^{\underline{\alpha}}$ to have the same form in all local reference frames. Both systems $\gamma^\alpha$ and $\gamma^{\underline{\alpha}}$ can be used to construct a complete system of 4x4 matrices. An example of a complete system can be the system $E, \gamma^\alpha, S^{\alpha\beta} = \frac{1}{2}\left(\gamma^\alpha\gamma^\beta - \gamma^\beta\gamma^\alpha\right)$, and $\gamma_5 = \gamma^0\gamma^1\gamma^2\gamma^3$ and $\gamma_5\gamma^\alpha$.

Any set of Dirac matrices can be used for several discrete automorphisms. We restrict ourselves to the automorphism $\gamma^\alpha \to \left(\gamma^\alpha\right)^+ = D\gamma^\alpha D^{-1}$; $D$ is called the Hermitizing matrix. The covariant derivative of the bispinor in expression (1) is

$$\nabla_\alpha \psi = \frac{\partial \psi}{\partial x^\alpha} + \Phi_\alpha \psi. \tag{4}$$

To determine the bispinor connectivities $\Phi_\alpha$ in (4), we must choose a certain system of tetrad vectors $H^\mu_{\underline{\alpha}}$ satisfying the relations $H^\mu_{\underline{\alpha}} H^\nu_{\underline{\beta}} g_{\mu\nu} = \eta_{\underline{\alpha}\underline{\beta}}$. In addition to the tetrad vectors $H^\mu_{\underline{\alpha}}$, we can introduce three other systems of tetrad vectors $H_{\underline{\alpha}\mu}, H^{\underline{\alpha}\mu}$, and $H^{\underline{\alpha}}_\mu$, which differ from $H^\mu_{\underline{\alpha}}$ by the positions of the world and local (underlined) indices. The world indices can be raised and lowered by the metric tensor $g_{\mu\nu}$ and by the inverse tensor $g^{\mu\nu}$, and the local indices can be raised and lowered by the tensors $\eta_{\underline{\alpha}\underline{\beta}}$ and $\eta^{\underline{\alpha}\underline{\beta}}$.

The bispinor connectivities can be defined by the Christoffel derivatives of tetrad vectors:

$$\Phi_\alpha = -\frac{1}{4} H^{\underline{\varepsilon}}_\mu H_{\nu\underline{\varepsilon};\alpha} S^{\mu\nu} = \frac{1}{4} H^{\underline{\varepsilon}}_\mu H_{\underline{\nu}\underline{\varepsilon};\alpha} S^{\underline{\mu}\underline{\nu}}. \tag{5}$$

The relationship between $\gamma^\alpha$ and $\gamma^{\underline{\alpha}}$ is given by the equality

$$\gamma^\alpha = H^\alpha_{\underline{\beta}} \gamma^{\underline{\beta}}. \tag{6}$$

Under a coordinate transformations

$$\{x^\alpha\} \to \{x'^\alpha\}, \tag{7}$$

we have the relations

$$\gamma'^\alpha = \frac{\partial x'^\alpha}{\partial x^\beta} \gamma^\beta, \quad \Phi'_\alpha = \frac{\partial x'^\beta}{\partial x^\alpha} \Phi_\beta. \tag{8}$$

The form of the wave functions of Dirac equation (1) is unchanged under transformation (7) except the corresponding change of variables.

Two arbitrary systems of tetrad vectors in the same space-time are related by the Lorentz transformation $L(x)$

$$\tilde{H}^\mu_{\underline{\alpha}}(x) = \Lambda^{\underline{\beta}}_{\underline{\alpha}}(x) H^\mu_{\underline{\beta}}(x). \tag{9}$$

The quantities $\Lambda^{\underline{\beta}}_{\underline{\alpha}}(x)$ satisfy the relations

$$\Lambda^{\mu}_{\underline{\alpha}}(x)\Lambda^{\nu}_{\underline{\beta}}(x)\eta^{\underline{\alpha\beta}} = \eta^{\mu\nu}, \quad \Lambda^{\mu}_{\underline{\alpha}}(x)\Lambda^{\nu}_{\underline{\beta}}(x)\eta_{\mu\nu} = \eta_{\underline{\alpha\beta}}. \tag{10}$$

The Dirac currents of particles are conserved under Lorentz transformations.

The introduced mathematical tool provides the covariance of Dirac equation (1) both under coordinate transformations (7) and in passing from one system of tetrad vectors to another (9).

### 2.2. Hamiltonians of Dirac equation

The Dirac equation in the Hamiltonian form is

$$i\frac{\partial\psi}{\partial t} = H\psi, \tag{11}$$

where $t = x^0$ in the left-hand side of (11) and $H$ is the Hamiltonian operator.

By virtue of expression (4) and the equality $\gamma^0\gamma^0 = g^{00}$, from Eq. (1), we can derive the Hamiltonian

$$H = \frac{m}{g^{00}}\gamma^0 - \frac{1}{g^{00}}i\gamma^0\gamma^k\frac{\partial}{\partial x^k} - i\Phi_0 - \frac{1}{g^{00}}i\gamma^0\gamma^k\Phi_k. \tag{12}$$

It was shown in [23] that in the same space-time, we can pass from any system of tetrad vectors $\{H^{\mu}_{\underline{\alpha}}(x)\}$ to the system of tetrad vectors $\{\tilde{H}^{\mu}_{\underline{\alpha}}(x)\}$ in the Schwinger gauge [24] using the Lorentz transformation $L(x)$.

For the system $\{\tilde{H}^{\mu}_{\underline{\alpha}}(x)\}$, we have

$$\tilde{H}^0_{\underline{0}} = \sqrt{g^{00}}, \quad \tilde{H}^k_{\underline{0}} = -\frac{g^{0k}}{\sqrt{g^{00}}}, \quad \tilde{H}^0_{\underline{k}} = 0. \tag{13}$$

Any spatial tetrads satisfying the relations

$$\tilde{H}^m_{\underline{k}}\tilde{H}^n_{\underline{k}} = f^{mn}, \quad f^{mn} = g^{mn} + \frac{g^{0m}g^{0n}}{g^{00}}, \quad f^{mn}g_{nk} = \delta^m_k \tag{14}$$

can be used as the tetrad vectors $\tilde{H}^n_{\underline{m}}$. The Lorentz transformation matrix is

$$L(x) = R\exp\left\{\frac{\theta}{2}\frac{\tilde{H}^{\mu}_{\underline{0}}H^{\nu}_{0}S_{\mu\nu}}{\sqrt{\left(\tilde{H}^{\varepsilon}_{\underline{0}}H_{0\varepsilon}\right)^2 - 1}}\right\}. \tag{15}$$

Here, $R$ represents a spatial rotation matrix commuting with $\gamma^{\underline{0}}$. The second factor is the transformation of the hyperbolic rotation (boost) through the angle $\theta$ defined by the relation

$$\text{th}\frac{\theta}{2} = \sqrt{\frac{\left(\tilde{H}^{\varepsilon}_{\underline{0}}H_{0\varepsilon}\right)+1}{\left(\tilde{H}^{\varepsilon}_{\underline{0}}H_{0\varepsilon}\right)-1}}. \tag{16}$$

The matrix $L$ transforms $\gamma^0(x)$ into the form $L\gamma^0 L^{-1} = \sqrt{g^{00}}\,\underline{\gamma}^{0}$.

Using some freedom to choose the spatial tetrads $\tilde{H}^n_{\underline{m}}$, which is defined by relations (14), passing from Hamiltonians (12) with the tetrad vectors $\{H^\mu_{\underline{\alpha}}(x)\}$ to Hamiltonians $\tilde{H}$ with the system of tetrad vectors $\{\tilde{H}^\mu_{\underline{\alpha}}(x)\}$ in the Schwinger gauge, and using different sets of $\tilde{H}^n_{\underline{m}}$, we can obtain expressions that do not coincide with each other. In fact, these Hamiltonians are physically equivalent because they are related by the unitary matrices of spatial rotations [25].

### 2.3. Hermiticity conditions for Hamiltonians and wave functions

In was shown in [23] that stationary Dirac Hamiltonians in an external gravitational field are pseudo-Hermitian and satisfy the condition of pseudo-Hermitian quantum mechanics [26] - [28]

$$H^+ = \rho H \rho^{-1}. \tag{17}$$

The operator $\rho$ in (17) is the Parker weight operator [29]: $\rho = \sqrt{-g}\,\gamma^0\underline{\gamma}^0$, where $g$ is the determinant of the metric $g_{\mu\nu}$. For the tetrad vectors in the Schwinger gauge, $\rho = \sqrt{-g}\sqrt{g^{00}}$. The scalar product of wave functions with $\rho$ is

$$(\varphi,\psi) = \int \varphi^+(x)\rho(x)\psi(x)d^3x. \tag{18}$$

The general Hermitician condition for Dirac Hamiltonians in an external gravitational field $(\varphi, H\psi) = (H\varphi, \psi)$ can be written as [30]

$$\oint ds_k\left(\sqrt{-g}\,j^k\right) + \int d^3x\sqrt{-g}\left[\psi^+\gamma^0\left(\gamma^0_{,0} + \begin{pmatrix} 0 \\ 0 & 0 \end{pmatrix}\gamma^0\right)\psi + \begin{pmatrix} k \\ k & 0 \end{pmatrix}j^0\right] = 0. \tag{19}$$

The current components in formula (19) are $j^\mu = \psi^+\gamma^0\gamma^\mu\psi$. For the time-independent Hamiltonians $\gamma^0_{,0} \equiv \partial\gamma^0/\partial x^0 = 0$, the Christoffel symbols $\begin{pmatrix} 0 \\ 0 & 0 \end{pmatrix}$ and $\begin{pmatrix} k \\ k & 0 \end{pmatrix}$ for the centrally symmetric fields are zero, and condition (19) can be written as

$$\oint ds_k\left(\sqrt{-g}\,j^k\right) = 0. \tag{20}$$

If there is an operator $\eta$ satisfying the relation

$$\left(\frac{g_G}{g}\right)^{1/2}\rho = \eta^+\eta, \tag{21}$$

then the Hamiltonian

$$H_\eta = \eta H \eta^{-1} \tag{22}$$

is self-conjugate, $H_\eta^+ = H_\eta$, and the scalar product (18) becomes flat (without the weight factor $\rho(x)$). In this case,

$$\psi_\eta(x) = \eta \psi(x). \tag{23}$$

We introduce the notation $g_G = g/g_C$ [25] in relation (21), where $g_C$ is the determinant arising if we write the volume element in the curvilinear coordinates of the flat background space ($g_C = 1$ for Cartesian coordinates, $g_C = r^2$ for cylindrical coordinates, $g_C = r^4 \sin^2\theta$ for spherical coordinates, and so on).

It was shown in [25] that the Hamiltonian $H_\eta$ given by (22) for the considered metrics of centrally symmetric gravitational fields can be obtained without direct calculations of bispinor connectivities (5) from the expression

$$H_\eta = \frac{1}{2}\left(\tilde{H}_{red} + \tilde{H}_{red}^+\right), \tag{24}$$

where $\tilde{H}_{red}$ is the part of initial Hamiltonian (12) with tetrads in the Schwinger gauge without terms with bispinor connectivities $\tilde{\Phi}_0$ and $\tilde{\Phi}_k$.

**2.4. Separation of variables**

The Dirac equation with Hamiltonian (24) and wave function (23)

$$i\frac{\partial \psi_\eta}{\partial t} = H_\eta \psi_\eta \tag{25}$$

admits separation of angular and radial variables in centrally symmetric gravitational fields.

Hamiltonians (24) for the static and stationary metrics are independent of time. To separate the variables, we represent the bispinor $\psi_\eta(t,r,\theta,\psi)$ in the form

$$\psi_\eta(t,r,\theta,\psi) = \begin{pmatrix} F(r)\xi(\theta) \\ -iG(r)\sigma^3\xi(\theta) \end{pmatrix} e^{-iEt} e^{im_\varphi \varphi} \tag{26}$$

and use the Brill–Wheeler equation [31]

$$\left[-\sigma^2\left(\frac{\partial}{\partial \theta} + \frac{1}{2}\mathrm{ctg}\,\theta\right) + i\sigma^1 m_\varphi \frac{1}{\sin\theta}\right]\xi(\theta) = i\kappa\xi(\theta). \tag{27}$$

For convenience in using Eq. (27), we perform an equivalent change of the local matrices in Hamiltonian (24),

$$\gamma^1 \to \gamma^3,\ \gamma^3 \to \gamma^2,\ \gamma^2 \to \gamma^1. \tag{28}$$

In Eqs. (26) and (27), $\xi(\theta)$ are the spherical harmonics for half-spin, $\sigma^k$ are the two-dimentional Pauli matrices, $E$ is the Dirac particle energy, $m_\varphi = -j, -j+1, \ldots,$ $j$ is the azimuthal component of the angular momentum $j$, $\kappa$ is the quantum number of the Dirac equation $\left(\kappa = \mp 1, \mp 2... = -(l+1),\ j = l + 1/2,\ \kappa = l,\ j = l - 1/2\right)$, and $j, l$ are the quantum numbers of the total angular and orbital momenta of the Dirac particle. The spinor $\xi(\theta)$ can be presented as in [32]

$$\xi(\theta) = \begin{pmatrix} _{-1/2}Y_{jm_\varphi}(\theta) \\ _{1/2}Y_{jm_\varphi}(\theta) \end{pmatrix} = (-1)^{m_\varphi + 1/2} \sqrt{\frac{1}{4\pi} \frac{(j - m_\varphi)!}{(j + m_\varphi)!}} \begin{pmatrix} \cos\theta/2 & \sin\theta/2 \\ -\sin\theta/2 & \cos\theta/2 \end{pmatrix} \times \\ \times \begin{pmatrix} (\kappa - m_\varphi + 1/2)\ P_l^{m_\varphi - 1/2}(\theta) \\ P_l^{m_\varphi + 1/2}(\theta) \end{pmatrix}. \tag{29}$$

The expression after the square root in (29), is a two-dimensional matrix, and $P_l^{m_\varphi \pm 1/2}(\theta)$ are associated Legendre polynomials.

As a result of separating the variables, the equations for the radial functions $F(\rho)$ and $G(\rho)$ can be written as

$$\frac{dF(\rho)}{d\rho} = A(\rho)F(\rho) + B(\rho)G(\rho),\quad \frac{dG(\rho)}{d\rho} = C(\rho)F(\rho) + D(\rho)G(\rho), \tag{30}$$

where $A(\rho), B(\rho), C(\rho)$ and $D(\rho)$ are determined by the form of Hamiltonian (24) for a particular metric.

In (30) and hereafter, we use the dimensionless variables $\rho = r/l_C$, $\varepsilon = E/m$, and $\alpha = r_0/2l_C = GMm/(\hbar c) = Mm/M_P^2$. Here, $l_C = \hbar/mc$ is the Compton wavelength for the Dirac particle, and $M_P = \sqrt{\hbar c/G} = 2.2 \cdot 10^{-5}$ g $(1.2 \cdot 10^{19}$ GeV) is the Planck mass.

**2.5 Self-conjugate second-order equation with a spinor wave function [33]**

We first square Dirac equation (25). For the static and stationary metrics, we have $i\,\partial\psi_\eta/\partial t = \varepsilon\psi_\eta$ (see representation (26)). We can then write the second-order equation as

$$(\varepsilon + H_\eta)(\varepsilon - H_\eta)\psi_\eta = 0. \tag{31}$$

We next pass in (31) from the equation for the bispinor $\psi_\eta$ to the equations for the upper and lower spinors in representation (26). For this, we must use the spinors relations in Eq. (25).

For the second-order equation with spinor wave functions to be self-conjugate, we must perform the nonunitary similarity transformations. After the variable separation procedure, the nonunitary similarity transformation for the upper spinor in (26) is

$$\Phi(\rho) = gF(\rho), \tag{32}$$

$$g(\rho) = \exp\left[\frac{1}{2}\int A_F(\rho')d\rho'\right], \tag{33}$$

$$A_F(\rho) = -\frac{1}{B}\frac{dB}{d\rho} - A - D, \tag{34}$$

where $A(\rho), B(\rho)$ and $D(\rho)$ are defined by Eqs. (30). As a result, the function $\Phi(\rho)$ satisfies the self-conjugate relativistic second-order equation of the Schrödinger type with the effective potential $U_{eff}$

$$\frac{d^2\Phi}{d\rho^2} + 2(E_{Schr} - U_{eff})\Phi = 0, \tag{35}$$

$$E_{Schr} = \frac{1}{2}(\varepsilon^2 - 1), \tag{36}$$

$$U_{eff} = -\frac{1}{4}\frac{1}{B}\frac{d^2B}{d\rho^2} + \frac{3}{8}\left(\frac{1}{B}\frac{dB}{d\rho}\right)^2 - \frac{1}{4}(A-D)\frac{1}{B}\frac{dB}{d\rho} + \frac{1}{4}\frac{d}{d\rho}(A-D) +$$
$$+ \frac{1}{8}(A-D)^2 + \frac{1}{2}BC + E_{Schr}. \tag{37}$$

The notation $E_{Schr}$ in (35) - (37) is introduced for conveniences. On the one hand, it provides the Schrödinger equation form for (35). On the other hand, effective potential (37) for $\varepsilon = 0$ has the classical asymptotic $U_{eff}\big|_{\rho\to\infty} = \alpha/\rho$.

**2.6. Road map for a quantum mechanical analysis of the equivalence of centrally symmetric solutions of GR equations**

As a base metric, we consider the Schwarzschild solution in the coordinates $(t, r, \theta, \varphi)$. All other centrally symmetric solutions to the GR equations can be obtained using appropriate coordinate transformations of the base metric.

For each metric, we obtain self-conjugate Dirac Hamiltonians with a flat scalar product of the wave functions by two methods. Using the first method, we obtain the Dirac Hamiltonians with tetrads in the Schwinger gauge (13), (14).

With the second method, we obtain self-conjugate Hamiltonians (in the $\eta$-representation with tetrads (13) and (14)) for the transformed metrics in two steps. The first step is to transform the basic self-conjugate Schwarzschild Hamiltonian to the coordinates of the transformed metric according to expressions (7) and (8), i.e., in this case the transformed tetrads are

$$H'^{\alpha}_{\underline{\beta}} = \frac{\partial x'^{\alpha}}{\partial x^{\mu}} H^{\mu}_{\underline{\beta}}. \tag{38}$$

The form of the wave function in (11) does not change apart from the change of variables. If the radial wave functions in (26) are real before transformation (7), then they remain real after transformations (7).

Next, if necessary, we perform the second step, which is Lorentz transformation (9), (10), (15), (16) to transform the Hamiltonian obtained in the coordinates of the transformed metric to the tetrads in the Schwinger gauge. In the second step, the currents of the Dirac particles do not change, but the form of the wave functions does change, and real radial functions can become complex.

In the final step of the transformation, we control the domain of the wave functions, currents of the Dirac particles, the Hermiticity of the Hamiltonian, and the possibility of the existence of stationary bound states of half-spin particles in using the second-order equation.

### 3. Schwarzschild metric in coordinates $(t, r, \theta, \varphi)$

The square of the interval is defined as

$$ds^2 = f_S dt^2 - \frac{dr^2}{f_S} - r^2 \left( d\theta^2 + \sin^2\theta d\varphi^2 \right), \tag{39}$$

where $f_S = 1 - r_0/r$. The timelike Killing vector has the form $\eta^\alpha = (1, 0, 0, 0)$.

### 3.1 Dirac equation

The nonzero tetrads in the Schwinger gauge $\tilde{H}^\mu_{\underline{\alpha}}$ are

$$\tilde{H}^0_{\underline{0}} = \frac{1}{\sqrt{f_S}}, \quad \tilde{H}^1_{\underline{1}} = \sqrt{f_S}, \quad \tilde{H}^2_{\underline{2}} = \frac{1}{r}, \quad \tilde{H}^3_{\underline{3}} = \frac{1}{r\sin\theta}. \tag{40}$$

According to relation (6), the matrices $\tilde{\gamma}^\alpha$ are equal to $\tilde{\gamma}^0 = \gamma^{\underline{0}}/\sqrt{f_S}$, $\tilde{\gamma}^1 = \sqrt{f_S}\gamma^{\underline{1}}$, $\tilde{\gamma}^2 = \gamma^{\underline{2}}/r$, and $\tilde{\gamma}^3 = \gamma^{\underline{3}}/(r\sin\theta)$. The self-conjugate Dirac Hamiltonian with tetrads (40) can be written as (see [23])

$$H_\eta = \eta \tilde{H}_S \eta^{-1} = \sqrt{f_S} m \gamma^{\underline{0}} - i\sqrt{f_S}\gamma^{\underline{0}} \left\{ \gamma^{\underline{1}} \sqrt{f_S}\left(\frac{\partial}{\partial r} + \frac{1}{r}\right) + \gamma^{\underline{2}} \frac{1}{r}\left(\frac{\partial}{\partial \theta} + \frac{1}{2}\operatorname{ctg}\theta\right) + \right.$$
$$\left. + \gamma^{\underline{3}} \frac{1}{r\sin\theta} \frac{\partial}{\partial \varphi} \right\} - \frac{i}{2} \frac{\partial f_S}{\partial r} \gamma^{\underline{0}}\gamma^{\underline{1}}. \tag{41}$$

In (41), $\tilde{H}_S$ is the Dirac Hamiltonian with tetrads (40).

After separation of the variables, by virtue of representation (26) and change (28), the equations for the radial functions become

$$f_S \frac{dF}{d\rho} + \left(\frac{1+\kappa\sqrt{f_S}}{\rho} - \frac{\alpha}{\rho^2}\right)F - \left(\varepsilon + \sqrt{f_S}\right)G = 0,$$
$$f_S \frac{dG}{d\rho} + \left(\frac{1-\kappa\sqrt{f_S}}{\rho} - \frac{\alpha}{\rho^2}\right)G + \left(\varepsilon - \sqrt{f_S}\right)F = 0. \tag{42}$$

Below, we consider the real radial functions $F(\rho) = F^*(\rho)$ and $G(\rho) = G^*(\rho)$. It follows from Eqs. (42) in this case that

$$\sqrt{f_S} \text{ is a positive real number} \tag{43}$$

and $f_S = 1 - r_0/r > 0$. The domain of the real radial functions is the region $r > r_0$. The transformation operator $\eta$ in (21) is $\eta = f_S^{-1/4}$. The components of the Dirac particle current $j^\mu = \psi_\eta^+ (\eta^{-1})^+ (\gamma^0 \tilde{\gamma}^\mu)(\eta^{-1})\psi_\eta$ are

$$j^0 = \psi_\eta^+ \psi_\eta = \left(F^*(\rho)F(\rho) + G^*(\rho)G(\rho)\right)\xi^+(\theta)\xi(\theta), \tag{44}$$

$$j^\rho = \psi_\eta^+ f_S \gamma^0 \gamma^3 \psi_\eta = -if_S\left(F^*(\rho)G(\rho) - F(\rho)G^*(\rho)\right)\xi^+(\theta)\xi(\theta), \tag{45}$$

$$j^\theta = \psi_\eta^+ \frac{\sqrt{f_S}}{\rho}\gamma^0\gamma^1\psi_\eta = -\frac{\sqrt{f_S}}{\rho}\left(F^*(\rho)G(\rho) + F(\rho)G^*(\rho)\right)\xi^+(\theta)\sigma^2\xi(\theta), \tag{46}$$

$$j^\varphi = \psi_\eta^+ \frac{\sqrt{f_S}}{\rho\sin\theta}\gamma^0\gamma^2\psi_\eta = \frac{\sqrt{f_S}}{\rho\sin\theta}\left(F^*(\rho)G(\rho) + F(\rho)G^*(\rho)\right)\xi^+(\theta)\sigma^1\xi(\theta). \tag{47}$$

We perform change (28) in formulas (44) - (47).

The radial density of the current $j^\rho$ given by (45) for real radial functions $(F^* = F, G^* = G)$ vanishes in the whole domain $(2\alpha, \infty)$. The current density $j^\theta$ given by (46) vanishes because $\xi^+(\theta)\sigma^2\xi(\theta) = 0$ (see expression (29)). The current density $j^\varphi$ given by (47) is nonzero in the domain $(2\alpha, \infty)$.

Condition (20) for the real radial functions becomes

$$\oint ds_\varphi \left(\sqrt{-g}\, j^\varphi\right) = 0. \tag{48}$$

Equality (48) holds in the spherically symmetric case, which proves the Hermiticity of Hamiltonian (41).

As $\rho \to \infty$, the leading asymptotic terms behave as (see, e.g. [32])

$$F\big|_{\rho\to\infty} = C_1\varphi_1(\rho)e^{-\rho\sqrt{1-\varepsilon^2}} + C_2\varphi_2(\rho)e^{\rho\sqrt{1-\varepsilon^2}},$$
$$G\big|_{\rho\to\infty} = \frac{\sqrt{1-\varepsilon}}{\sqrt{1+\varepsilon}}\left(-C_1\varphi_1(\rho)e^{-\rho\sqrt{1-\varepsilon^2}} + C_2\varphi_2(\rho)e^{\rho\sqrt{1-\varepsilon^2}}\right). \tag{49}$$

In formulas (49), $\varphi_1(\rho)$ and $\varphi_2(\rho)$ are power functions of $\rho$, and $C_1$ and $C_2$ are integration constants. To ensure a finitary motion of Dirac particles, we must use only the exponentially decaying solutions (49), i.e., $C_2 = 0$ in this case.

If we write $F(\rho)$ and $G(\rho)$ as $\rho \to 2\alpha$ in the forms

$$F\big|_{\rho \to 2\alpha} = (\rho - 2\alpha)^s \sum_{k=0}^{\infty} f_k (\rho - 2\alpha)^k, \quad G\big|_{\rho \to 2\alpha} = (\rho - 2\alpha)^s \sum_{k=0}^{\infty} g_k (\rho - 2\alpha)^k, \tag{50}$$

then the indicial equation for system (42) leads to the solution

$$s = -\frac{1}{2} \pm i 2\alpha\varepsilon. \tag{51}$$

For the real radial functions, we can write the asymptotic forms by virtue of (50) and (51) as

$$F\big|_{\rho \to 2\alpha} = \frac{L}{\sqrt{\rho - 2\alpha}} \sin\big(2\alpha\varepsilon \ln(\rho - 2\alpha) + \varphi\big),$$

$$G_{\rho \to 2\alpha} = \frac{L}{\sqrt{\rho - 2\alpha}} \cos\big(2\alpha\varepsilon \ln(\rho - 2\alpha) + \varphi\big). \tag{52}$$

In Eqs. (52), $L$ and $\varphi$ are integration constants. The functions $F$ and $G$ in expressions (42) are not square integrable as $\rho \to 2\alpha$. The form of the oscillating parts of $F$ and $G$ for $\varepsilon \neq 0$ justifies the realization of the regime of a particle "fall" on the event horizon [21], [22].

### 3.2 Self-conjugate second-order equations with a spinor wave function

To obtain square-integrable real radial functions in the domain $\rho \in (2\alpha, \infty)$, we pass from system of equations (42) to Schrödinger-type relativistic equation (35) for the function $\Phi(\rho)$ proportional to $F(\rho)$ (see formulas (32) - (37)). Using the Schwarzschild metric in the coordinates $(t, r, \theta, \varphi)$, we write the expressions for $A(\rho), B(\rho), C(\rho)$, and $D(\rho)$ as (see formulas (30) and (42)

$$A(\rho) = -\frac{1}{f_S}\left(\frac{1 + \kappa\sqrt{f_S}}{\rho} - \frac{\alpha}{\rho^2}\right), \quad B(\rho) = \frac{1}{f_S}\left(\varepsilon + \sqrt{f_S}\right),$$

$$C(\rho) = -\frac{1}{f_S}\left(\varepsilon - \sqrt{f_S}\right), \quad D(\rho) = -\frac{1}{f_S}\left(\frac{1 - \kappa\sqrt{f_S}}{\rho} - \frac{\alpha}{\rho^2}\right). \tag{53}$$

The explicit form of the effective potential $U_{\mathit{eff}}$ in (37) for the Schwarzschild field in the coordinates $(t, r, \theta, \varphi)$ is given in Appendix A.

### 3.3 Degenerate stationary bound states of fermions with zero energy $\varepsilon = 0$

We see from asymptotic formulas (52) that the regime of a particle "fall" on the event horizon is absent for only the fermion energy value $\varepsilon = 0$. For $\varepsilon = 0$, the asymptotic form of effective potential (37) near the event horizon is

$$U_{eff}(\varepsilon = 0)\big|_{\rho \to 2\alpha} = -\frac{3}{32}\frac{1}{(\rho - 2\alpha)^2}. \tag{54}$$

The coefficient $3/32 < 1/8$, which according to [21] indicates the absence of the regime of a particle "fall" on the event horizon and the possibility of the existence of a stationary bound state with $\varepsilon = 0$.

The transformation $g(\rho)$ given by (33) can be written asymptotically es

$$g\big|_{\rho \to \infty} = \rho, \quad g\big|_{\rho \to 2\alpha} = (\rho - 2\alpha)^{3/4}. \tag{55}$$

As a result, in view of (49) and (52), the asymptotic expressions for the transformed function $\Phi(\rho)$ given by (32) satisfying the second-order equation (35) are

$$\Phi\big|_{\rho \to \infty} = C_1 \varphi_1(\rho) \rho e^{-\rho\sqrt{1-\varepsilon^2}}, \quad \Phi\big|_{\rho \to 2\alpha} = L(\rho - 2\alpha)^{1/4}. \tag{56}$$

Equalities (56) indicate that the transformed function $\Phi(\rho)$ is square integrable. The fermion wave function vanishes on the event horizon.

The numerical calculations of the solution of second-order equation (35) confirm the existence of a degenerate solution with $\varepsilon = 0$ [20]. The fermions in the bound states with $\varepsilon = 0$ an overwhelming probability are localized in a neighborhood near the outside of the event horizon inside the intervals from zero to fractions of the fermion Compton wavelength, depending on the value of the gravitational coupling constant $\alpha$ and the quantities $j$ and $l$.

### 3.4 Coordinate transformations of the Dirac equation

In what follows, we consider three types of transformations:

1. For the static Schwarzschild metrics in isotropic coordinates, $\eta^\alpha(1,0,0,0)$, we transform the spherical coordinates of the original Schwarzschild metric $(t, r, \theta, \varphi)$ to the new coordinates $(t, R, \theta, \varphi)$, where the radial coordinate $R = \varphi(r)$ is a function of the original coordinate $r$. In this case, representation (26) holds in the new variables, and the Dirac equation there is subjected to the standard change of variables

$$Ee^{-iEt}\tilde{\psi}_{is}(\mathbf{R}) = \tilde{H}_{is}(\mathbf{R})e^{-iEt}\tilde{\psi}_{is}(\mathbf{R}). \tag{57}$$

All the physical consequences found for the Dirac equation in Sec. 3.1 and for the second-order equation in Sec. 3.2 hold for transformed equation (57). This conclusion also relates to the existence of a degenerate bound state of fermions with $E=0$.

The form of Eq. (57) in the $\eta$-representation $\left(\psi_\eta = \eta\tilde{\psi}, H_\eta = \eta\tilde{H}\eta^{-1}\right)$ differs from the original equation with Hamiltonian (41). But the $\eta$-transformation does not affect the energy of the Dirac particle and does not contribute new physical effects.

2. For the stationary EF and PG metrics, $\eta^\alpha(1,0,0,0)$, the time coordinate is subjected to a transformation. For the EF metric $(t,r,\theta,\varphi) \to (T_{EF}, r, \theta, \varphi)$

$$dT_{EF} = dt + \frac{r_0}{r}\frac{dr}{f_S}, \quad T_{EF} = t + \varphi_{EF}(r) = t + \int \frac{r_0}{r}\frac{dr}{f_S}.$$

For the PG metric $(t,r,\theta,\varphi) \to (T_{PG}, r, \theta, \varphi)$

$$dT_{PG} = dt + \sqrt{\frac{r_0}{r}}\frac{1}{f_S}dr, \quad T_{PG} = t + \varphi_{PG}(r) = t + \int \sqrt{\frac{r_0}{r}}\frac{dr}{f_S}.$$

We can write representation (26) and the Dirac equation for both metrics with the new time variable $T(t,r)$ as

$$Ee^{-iE(T_{EF} - \varphi_{EF}(r))}\tilde{\psi}_{EF}(\mathbf{r}) = \tilde{H}_{EF}(\mathbf{r})e^{-iE(T_{EF} - \varphi_{EF}(r))}\tilde{\psi}_{EF}(\mathbf{r}),$$

$$Ee^{-iE(T_{PG} - \varphi_{PG}(r))}\tilde{\psi}_{PG}(\mathbf{r}) = \tilde{H}_{PG}(\mathbf{r})e^{-iE(T_{PG} - \varphi_{PG}(r))}\tilde{\psi}_{PG}(\mathbf{r}),$$

$$E\tilde{\psi}_{EF}(\mathbf{r}) = e^{-iE\varphi_{EF}(r)}\tilde{H}_{EF}(\mathbf{r})e^{+iE\varphi_{EF}(r)}\tilde{\psi}_{EF}(\mathbf{r}), \tag{58}$$

$$E\tilde{\psi}_{PG}(\mathbf{r}) = e^{-iE\varphi_{PG}(r)}\tilde{H}_{PG}(\mathbf{r})e^{+iE\varphi_{PG}(r)}\tilde{\psi}_{PG}(\mathbf{r}), \tag{59}$$

where $\tilde{\psi}_S(\mathbf{r}) = \tilde{\psi}_{EF}(\mathbf{r}) = \tilde{\psi}_{PG}(\mathbf{r})$.

3. For the nonstationary LF and KSz metrics, $\eta^\alpha_{LF} = (1,1,0,0)$ and $\eta^\alpha_{KS} = (\eta^0, \eta^1, 0, 0)$,

Where

$$\eta^0 = \frac{1}{2r_0}\sqrt{\frac{r}{r_0}}\sqrt{f_S}\exp\left[\frac{r}{2r_0}\right]\text{ch}\frac{t}{2r_0}, \quad \eta^1 = \frac{1}{2r_0}\sqrt{\frac{r}{r_0}}\sqrt{f_S}\exp\left[\frac{r}{2r_0}\right]\text{sh}\frac{t}{2r_0}.$$

In this case, both the time and radial coordinates are transformed. For the LF metric, $(t,r,\theta,\varphi) \to (T_{LF}, R_{LF}, \theta, \varphi)$

$$dT_{LF} = dt + \frac{\sqrt{r_0/r}}{f_S}dr; \quad dR_{LF} = dt + \frac{dr}{f_S\sqrt{r_0/r}}.$$

For the KSz metric, $(t,r,\theta,\varphi) \to (v,u,\theta,\varphi)$

$$v = \sqrt{\frac{r}{r_0}} \sqrt{f_S} \exp\frac{r}{2r_0} \text{sh}\left(\frac{t}{2r_0}\right), \quad u = \sqrt{\frac{r}{r_0}} \sqrt{f_S} \exp\frac{r}{2r_0} \text{ch}\left(\frac{t}{2r_0}\right).$$

The Dirac Hamiltonian in both LF and KSz fields depends on the time coordinate. Therefore, we cannot represent the wave function in a form of proportional to $e^{-iET}$.

### 4. Schwarzschild metric in isotropic coordinates

The square of the interval is defined as

$$ds^2 = V^2(R)dt^2 - W^2(R)\left[dR^2 + R^2\left(d\theta^2 + \sin^2\theta d\varphi^2\right)\right], \tag{60}$$

where $V(R) = (1 - r_0/4R)/(1 + r_0/4R)$ and $W(R) = (1 + r_0/4R)^2$. The quantities $(-g)$, $g_G$ and $\eta$ have the forms $-g = V^2 W^6 R^4 \sin^2\theta$, $g_G = V^2 W^6$, and $\eta = (g_G)^{1/4}(g^{00})^{1/4} = W^{3/2}$. The coordinate transformation, nonzero tetrads $\tilde{H}^\mu_{\underline{\alpha}}$ and matrices $\tilde{\gamma}^\alpha$ are given in Appendix B (see formulas (B.1) – (B.4)).

#### 4.1 Dirac equation

The self-conjugate Hamiltonian in the $\eta$-representation with tetrads (B.3) is

$$\begin{aligned}H_\eta &= \frac{1 - r_0/4R}{1 + r_0/4R} m\gamma^0 - i\frac{1 - r_0/4R}{(1 + r_0/4R)^3} \gamma^0\left[\gamma^1\left(\frac{\partial}{\partial R} + \frac{1}{R}\right)\right] + \\ &\quad + \gamma^2 \frac{1}{R}\left(\frac{\partial}{\partial \theta} + \frac{1}{2}\text{ctg}\theta\right) + \gamma^3 \frac{1}{R\sin\theta}\frac{\partial}{\partial \theta}\bigg] - \frac{i}{2}\gamma^0\gamma^1 \frac{\partial}{\partial R}\frac{1 - r_0/4R}{(1 + r_0/4R)^3}.\end{aligned} \tag{61}$$

After separation of the variables, in view of representation (26) and change (28), the equations for the radial functions $F_{is}(\rho)$ and $G_{is}(\rho)$ in the dimensionless variables $\rho = R/l_C$, $\varepsilon = E/m$, and $\alpha = r_0/2l_C$ become

$$\begin{aligned}\frac{V}{W}\frac{dF_{is}}{d\rho} + \frac{V}{W}\frac{1+\kappa}{\rho}F_{is} + \frac{1}{2}\frac{d}{d\rho}\left(\frac{V}{W}\right)F_{is} - (\varepsilon + V)G_{is} &= 0, \\ \frac{V}{W}\frac{dG_{is}}{d\rho} + \frac{V}{W}\frac{1-\kappa}{\rho}G_{is} + \frac{1}{2}\frac{d}{d\rho}\left(\frac{V}{W}\right)G_{is} + (\varepsilon - V)F_{is} &= 0.\end{aligned} \tag{62}$$

To write the components of the current density, we use the tetrads $\tilde{H}^\mu_{\underline{\alpha}}$ in the Schwinger gauge given by (B.3): $j^\mu = \psi_\eta^+ (\eta^{-1})^+ \gamma^0 \tilde{\gamma}^\mu (\eta^{-1}) \psi_\eta$.

We next use equivalent change (28) and apply representation (28) with the radial functions $F_{is}(\rho), G_{is}(\rho)$. Then

$$j^0 = \frac{1}{VW^3}\left(F_{is}^*(\rho)F_{is}(\rho) + G_{is}^*(\rho)G_{is}(\rho)\right)\xi^+(\theta)\xi(\theta), \tag{63}$$

$$j^\rho = -i\frac{1}{W^4}\left(F_{is}*(\rho)G_{is}(\rho) - F_{is}(\rho)G_{is}*(\rho)\right)\xi^+(\theta)\xi(\theta), \tag{64}$$

$$j^\theta = -\frac{1}{RW^4}\left(F_{is}*(\rho)G_{is}(\rho) + F_{is}(\rho)G_{is}*(\rho)\right)\xi^+(\theta)\sigma^2\xi(\theta), \tag{65}$$

$$j^\varphi = \frac{1}{W^4 R\sin\theta}\left(F_{is}*(\rho)G_{is}(\rho) + F_{is}(\rho)G_{is}*(\rho)\right)\xi^+(\theta)\sigma^1\xi(\theta). \tag{66}$$

We now obtain Hamiltonian (61) by directly transforming basic Hamiltonian (41) with tetrads (40). Under coordinate transformation (B.2), tetrads (40) transform according to equality (38):

$$\left(H'^0_{\underline{0}}\right)_{is} = \frac{\partial t}{\partial t}\left(H^0_{\underline{0}}\right)_S = \frac{1}{\sqrt{f_S}} = \frac{1 + r_0/4R}{1 - r_0/4R}, \tag{67}$$

$$\left(H'^1_{\underline{1}}\right)_{is} = \frac{\partial R}{\partial r}\left(H^1_{\underline{1}}\right)_S = \frac{\sqrt{f_S}}{1 - r_0^2/16R^2} = \frac{1}{(1 + r_0/4R)^2}, \tag{68}$$

$$\left(H'^2_{\underline{2}}\right)_{is} = \left(H^2_{\underline{2}}\right)_S = \frac{1}{r} = \frac{1}{R(1 + r_0/4R)^2}, \tag{69}$$

$$\left(H'^3_{\underline{3}}\right)_{is} = \left(H^3_{\underline{3}}\right)_S = \frac{1}{r\sin\theta} = \frac{1}{R\sin\theta(1 + r_0/4R)^2}. \tag{70}$$

The transformed tetrads coincide with tetrads (B.3) in the Schwinger gauge for the Schwarzschild metric in the isotropic coordinates. According to formula (24), the Hamiltonian with tetrads (67) - (70) coincides with Hamiltonian (61).

We note that with the definition of the tetrad $\left(H'^0_{\underline{0}}\right)_{is}$ given by (67), the condition $R \geq r_0/4$ must be satisfied to preserve the positivity condition for $\sqrt{f_S}$ (see (43)), i.e., we necessarily obtain the domain for the wave functions of the Dirac equation with the Schwarzschild metric in isotropic coordinates $R > r_0/4$. With this domain, we aliminate the ambiguity noted in (B.2) in the definition of the coordinate $R$. If we use reals radial functions in Dirac equations (42) for the original Schwarzschild metric, then according to change (57) in Eqs. (62), the functions $F_{is}(\rho), G_{is}(\rho)$ are also real.

Below, we briefly give basic conclusions obtained in terms of the coordinates of transformed metric (60). In Sec. 3 above, we obtain analogous results in the coordinates of Schwarzschild metric $(t, r, \theta, \varphi)$. The difference is in changing the radius of the event horizon in transformed metric (60): $R = r_0/4$.

We thus formulate the basic conclusions:

1. For the real radial functions $\left(F_{is}* = F_{is}, \text{ and } G_{is}* = G_{is}\right)$, the radial density $j^\rho$ given

by (64) vanishes in the whole domain $(\alpha/2,\infty)$. The current density $j^\theta$ (65) is also vanishes.

2. Condition (20) holds, which proves the Hermiticity of the Hamiltonian (61).
3. As $\rho \to \alpha/2$, the determining equation for system (62) leads to the solution

$$s = -\frac{1}{2} \pm i4\alpha\varepsilon. \tag{71}$$

By virtue of (71), we can write the asymptotic formulas for the real radial functions as

$$\begin{aligned} F_{is}\big|_{\rho \to \frac{\alpha}{2}} &= \frac{C_3}{\sqrt{\rho - \alpha/2}} \sin\left(4\alpha\varepsilon \ln(\rho - \alpha/2) + \varphi_{is}\right), \\ G_{is}\big|_{\rho \to \frac{\alpha}{2}} &= \frac{C_3}{\sqrt{\rho - \alpha/2}} \cos\left(4\alpha\varepsilon \ln(\rho - \alpha/2) + \varphi_{is}\right), \end{aligned} \tag{72}$$

where $C_3$ and $\varphi_{is}$ are integration constants. The functions $F_{is}(\rho)$ and $G_{is}(\rho)$ in (62) are not square integrable as $\rho \to \alpha/2$. The form of the oscillating parts of $F_{is}(\rho)$ and $G_{is}(\rho)$ for $\varepsilon \neq 0$ in formulas (72) implies the realization of the regime of a particle "fall" on the event horizon [21], [22].

4. The expressions $A_{is}(\rho), B_{is}(\rho), C_{is}(\rho)$ and $D_{is}(\rho)$, which are needed to obtain a self-conjugate second-order equation with the spinor wave function, are (see formulas (30) and (62))

$$\begin{aligned} A_{is}(\rho) &= -\left(\frac{1+\kappa}{\rho} + \frac{W}{V}\frac{1}{2}\frac{d}{d\rho}\left(\frac{V}{W}\right)\right), \quad B_{is}(\rho) = \frac{W}{V}(\varepsilon + V), \\ C_{is}(\rho) &= -\frac{W}{V}(\varepsilon - V), \quad D_{is}(\rho) = -\left(\frac{1-\kappa}{\rho} + \frac{W}{V}\frac{1}{2}\frac{d}{d\rho}\left(\frac{V}{W}\right)\right). \end{aligned}$$

According to relations (32) - (37),

$$\Phi_{is}(\rho) = g_{is} F_{is}(\rho), \tag{73}$$

$$g_{is}(\rho) = \exp\frac{1}{2}\int A_F^{is}(\rho')d\rho', \tag{74}$$

$$A_F^{is}(\rho) = -\frac{1}{B_{is}}\frac{dB_{is}}{d\rho} - A_{is} - D_{is}. \tag{75}$$

Next,

$$\frac{d^2\Phi_{is}}{d\rho^2} + 2\left(E_{Schr} - U_{eff}^{is}\right)\Phi_{is} = 0, \tag{76}$$

$$E_{Schr} = \frac{1}{2}\left(\varepsilon^2 - 1\right), \tag{77}$$

$$U_{eff}^{is} = -\frac{1}{4}\frac{1}{B_{is}}\frac{d^2 B_{is}}{d\rho^2} + \frac{3}{8}\left(\frac{1}{B_{is}}\frac{dB_{is}}{d\rho}\right)^2 - \frac{1}{4}(A_{is} - D_{is})\frac{1}{B_{is}}\frac{dB_{is}}{d\rho} +$$
$$+\frac{1}{4}\frac{d}{d\rho}(A_{is} - D_{is}) + \frac{1}{8}(A_{is} - D_{is})^2 + \frac{1}{2}B_{is}C_{is} + E_{Schr.} \quad (78)$$

5. From asymptotic forms (72), we see that the regime of a particle "fall" on the event horizon is absent for only the fermion energy value $\varepsilon = 0$. For $\varepsilon = 0$, the asymptotic form of effective potential (78) near the event horizon is $U_{eff}^{is}(\varepsilon = 0)\big|_{\rho \to \alpha/2} = -(3/32)(1)/((\rho - \alpha/2)^2)$. The coefficient $3/32$ is less than $1/8$, which, on one hand, implies the absence of the regime of a particle "fall" on the event horizon and, on the other hand, implies possible existence of a stationary bound state with $\varepsilon = 0$. The transformation $g_{is}(\rho)$ given by (74) can be written asymptotically as $g_{is}(\rho)\big|_{\rho \to \infty} = \rho$, $g_{is}(\rho)\big|_{\rho \to \alpha/2} = (\rho - \alpha/2)^{3/4}$. By virtue of (49) and (72), the asymptotic forms of the function $\Phi(\rho)$ given by (73) satisfying second-order equation (76) are given by

$$\Phi_{is}\big|_{\rho \to \infty} = C_1 \varphi_1(\rho)\rho e^{-\rho\sqrt{1-\varepsilon^2}}, \quad (79)$$

$$\Phi_{is}\big|_{\rho \to \frac{\alpha}{2}} = C_3 \left(\rho - \frac{\alpha}{2}\right)^{1/4}. \quad (80)$$

In (79), we take into account that $R = r$ as $R \to \infty$ (see formula (B.2)). Equalities (79) and (80) imply that $\Phi_{is}(\rho)$ is square integrable. The fermion wave function vanishes on the event horizon. Just as for original Schwarzschild metric (39), numerical calculations of the solution of second-order equation (76) confirm the existence of the degenerate solution with $\varepsilon = 0$.

### 5. Eddington-Finkelstein and Painlevé-Gullstrand metrics

#### 5.1 Eddington-Finkelstein solution

The square of the interval is given by the expression

$$ds^2 = f_S dT^2 - 2\frac{r_0}{r}dTdr - \left(1 + \frac{r_0}{r}\right)dr^2 - r^2\left(d\theta^2 + \sin^2\theta d\varphi^2\right). \quad (81)$$

The quantities $(-g)$, $g_G$, and $\eta$ have the forms $-g = r^4 \sin^2\theta$, $g_G = 1$, and $\eta = (1 + r_0/r)^{1/4}$. The coordinate transformation, nonzero tetrads $\tilde{H}_{\underline{\alpha}}^{\mu}$ and matrices $\tilde{\gamma}^{\alpha}$ are given in Appendix B (see formulas (B.5) – (B.8)).

The self-conjugate Hamiltonian in the $\eta$- representation with tetrads (B.7) is [25]

$$H_\eta = \frac{m}{\sqrt{1+r_0/r}}\gamma^{\underline{0}} - i\gamma^{\underline{0}}\gamma^{\underline{1}}\frac{1}{1+r_0/r}\left(\frac{\partial}{\partial r}+\frac{1}{r}+\frac{r_0}{2r^2}\frac{1}{1+r_0/r}\right)-$$
$$-i\gamma^{\underline{0}}\gamma^{\underline{2}}\frac{1}{\sqrt{1+r_0/r}}\frac{1}{r}\left(\frac{\partial}{\partial\theta}+\frac{1}{2}\operatorname{ctg}\theta\right)-i\gamma^{\underline{0}}\gamma^{\underline{3}}\frac{1}{\sqrt{1+r_0/r}}\frac{1}{r\sin\theta}\frac{\partial}{\partial\varphi}+ \tag{82}$$
$$+i\frac{r_0}{r}\frac{1}{1+r_0/r}\left(\frac{\partial}{\partial r}+\frac{1}{r}-\frac{1}{2r(1+r_0/r)}\right).$$

We next obtain the Dirac Hamiltonian in the EF field by a direct transformation of basic Hamiltonian (41) with tetrads (40). Under coordinate transformation (B.6), the nonzero tetrads transformed according to (38) are

$$\left(H'^{0}_{\underline{0}}\right)_{EF} = \frac{\partial T}{\partial t}\left(H^{0}_{\underline{0}}\right)_S = \frac{1}{\sqrt{f_S}}; \tag{83}$$

$$\left(H'^{0}_{\underline{1}}\right)_{EF} = \frac{\partial T}{\partial r}\left(H^{1}_{\underline{1}}\right)_S = \frac{\frac{r_0}{r}}{\sqrt{f_S}}; \tag{84}$$

$$\left(H'^{1}_{\underline{1}}\right)_{EF} = \frac{\partial r}{\partial r}\left(H^{1}_{\underline{1}}\right)_S = \sqrt{f_S}; \tag{85}$$

$$\left(H'^{2}_{\underline{2}}\right)_{EF} = \left(H^{2}_{\underline{2}}\right)_S = \frac{1}{r}; \tag{86}$$

$$\left(H'^{3}_{\underline{3}}\right)_{EF} = \left(H^{3}_{\underline{3}}\right)_S = \frac{1}{r\sin\theta}. \tag{87}$$

As a result of transformations (B.6), the nonzero tetrad $\left(H'^{0}_{\underline{1}}\right)_{EF}$ given by (84) appears in addition to tetrads (40).

Coordinate transformation (B.6) does not change the form of the wave functions $\tilde{\psi}(\mathbf{r})$, i.e. $\tilde{\psi}_S(\mathbf{r})=\tilde{\psi}_{EF}(\mathbf{r})$. Equality (58) shows that for the solution $E=0$, the Hamiltonian for transformed metric (81) coincides with the Hamiltonian for original Schwarzschild metric (39) up to a matrix functional factor:

$$\tilde{H}_{EF}(\mathbf{r}) = P_{EF}(r)\tilde{H}_S(\mathbf{r}). \tag{88}$$

The definition of the Hamiltonian $\tilde{H}_{EF}(\mathbf{r})$ together with tetrads (83) - (87) and with calculation of bispinor connectivities (5) shows that we can write the factor $P_{EF}(r)$ as

$$P_{EF}(r) = \frac{1}{f_S(1+r_0/r)}\left(1+\frac{r_0}{r}\gamma^{\underline{0}}\gamma^{\underline{1}}\right). \tag{89}$$

By virtue of representation (26) with the radial functions $F_{EF}(\rho)$ and $G_{EF}(\rho)$ and change (28), the components of the current density for tetrads (83) - (87) are

$$j^0 = \frac{1}{\sqrt{1-\frac{4\alpha^2}{\rho^2}}} \psi_\eta^+ \psi_\eta = \frac{1}{\sqrt{1-\frac{4\alpha^2}{\rho}}} \Big[ \big( F_{EF}^*(\rho) F_{EF}(\rho) + G_{EF}^*(\rho) G_{EF}(\rho) \big) \xi^+(\theta) \xi(\theta) -$$

$$-i\frac{r_0}{r} \big( F_{EF}^*(\rho) G_{EF}(\rho) - F(\rho) G_{EF}^*(\rho) \big) \xi^+(\theta) \xi(\theta) \Big],$$

$$j^\rho = -i \left( \frac{1-2\alpha/\rho}{1+2\alpha/\rho} \right)^{1/2} \big( F_{EF}^*(\rho) G_{EF}(\rho) - F_{EF}(\rho) G_{EF}^*(\rho) \big) \xi^+(\theta) \xi(\theta),$$

$$j^\theta = -\frac{1}{\rho(1+2\alpha/\rho)^{1/2}} \big( F_{EF}^*(\rho) G_{EF}(\rho) + F_{EF}(\rho) G_{EF}^*(\rho) \big) \xi^+(\theta) \sigma^2 \xi(\theta),$$

$$j^\varphi = \frac{1}{\rho \sin\theta (1+2\alpha/\rho)^{1/2}} \big( F_{EF}^*(\rho) G_{EF}(\rho) + F_{EF}(\rho) G_{EF}^*(\rho) \big) \xi^+(\theta) \sigma^1 \xi(\theta).$$

Using the Lorentz transformation (9), (10), (15), and (16), we transform tetrads (83) - (87) to tetrads (B.7) in the Schwinger gauge. The nonzero quantities $\Lambda_{\underline{\alpha}}^{\underline{\beta}}(r)$ in (9) and (10) are

$$\Lambda_{\underline{0}}^{\underline{0}} = \Lambda_{\underline{1}}^{\underline{1}} = \frac{1}{\sqrt{f_S}\sqrt{1+r_0/r}}; \quad \Lambda_{\underline{0}}^{\underline{1}} = \Lambda_{\underline{1}}^{\underline{0}} = -\frac{r_0/r}{\sqrt{f_S}\sqrt{1+r_0/r}}. \tag{90}$$

After the transformation, Hamiltonian (88) has form (82). In the transformation process, the expression $\sqrt{f_S} = \sqrt{1-r_0/r}$ appears in formulas (83) - (87) and (90); it is real in basic metric (39) (see condition (43)). It hence follows that the domain of the wave functions of Hamiltonian (82) is again the region $r \in (r_0, \infty)$. The Hilbert causality condition $g_{00} > 0$ for EF metric (81) also leads to the domain $r > r_0$.

Coordinate transformation (B.6) preserves the realness of the radial functions $F_{EF}(\rho)$ and $G_{EF}(\rho)$ for tetrads (83) - (87). Therefore, the radial and $\theta$ components of the current density vanish, and the Hermiticity condition for Hamiltonian (19) also holds in the case of the EF metric.

According to formula (88), all conclusions in Sec. 3 for original Schwarzschild metric (39) also hold for the EF metric. In the EF gravitational field, there are degenerate stationary bound states of fermions with zero energy $E = 0$ and with square-integrable spinor wave functions satisfying the self-conjugate second-order equations.

It is interesting to answer the question whether our conclusions change if instead of Hamiltonian (88), we use self-conjugate Hamiltonian (82) in the $\eta$-representation with tetrads in the Schwinger gauge. Because the EF and PG metrics are similar, we answer this question in the next subsection.

## 5.2 Painlevé-Gullstrand solution

The square of the interval is given by the expression

$$ds^2 = f_S dT^2 - 2\sqrt{\frac{r_0}{r}} dT dr - dr^2 - r^2 \left( d\theta^2 + \sin^2\theta d\varphi^2 \right). \tag{91}$$

The quantities $(-g)$, $g_G$, and $\eta$ are defined by expressions $-g = r^4 \sin^2\theta$, $g_G = 1$, and $\eta = 1$. The coordinate transformation, nonzero tetrads $\tilde{H}^\mu_{\underline{\alpha}}$, and matrices $\tilde{\gamma}^\alpha$ are given in Appendix B (see formulas (B.9) – (B.12)).

The self-conjugate Hamiltonian in the $\eta$ - representation with tetrads (B.11) is [19], [25]

$$H_\eta = \gamma^{\underline{0}} m - i\gamma^{\underline{0}} \left\{ \gamma^{\underline{1}} \left( \frac{\partial}{\partial r} + \frac{1}{r} \right) + \gamma^{\underline{2}} \frac{1}{r} \left( \frac{\partial}{\partial \theta} + \frac{1}{2} \operatorname{ctg}\theta \right) + \right.$$
$$\left. + \gamma^{\underline{3}} \frac{1}{r \sin\theta} \frac{\partial}{\partial \varphi} \right\} + i \sqrt{\frac{r_0}{r}} \left( \frac{\partial}{\partial r} + \frac{3}{4} \frac{1}{r} \right). \tag{92}$$

We next obtain the Dirac Hamiltonian (92) by directly transforming basic Hamiltonian (41) with tetrads (40). Under coordinate transformation (B.10), the nonzero tetrads transformed according to expression (38) are

$$\left( H'^0_{\underline{0}} \right)_{PG} = \frac{\partial T}{\partial t} \left( H^0_{\underline{0}} \right)_S = \frac{1}{\sqrt{f_S}}, \tag{93}$$

$$\left( H'^0_{\underline{1}} \right)_{PG} = \frac{\partial T}{\partial r} \left( H^1_{\underline{1}} \right)_S = \frac{\sqrt{r_0/r}}{\sqrt{f_S}}, \tag{94}$$

$$\left( H'^1_{\underline{1}} \right)_{PG} = \frac{\partial r}{\partial r} \left( H^1_{\underline{1}} \right)_S = \sqrt{f_S}, \tag{95}$$

$$\left( H'^2_{\underline{2}} \right)_{PG} = \left( H^2_{\underline{2}} \right)_S = r^{-1}, \tag{96}$$

$$\left( H'^3_{\underline{3}} \right)_{PG} = \left( H^3_{\underline{3}} \right)_S = \left( r \sin\theta \right)^{-1}. \tag{97}$$

Compared with tetrads (40), an additional nonzero tetrad $\left( H'^0_{\underline{1}} \right)_{PG}$ (94) appeared as a result of transformation (B.10).

Coordinate transformation (B.10) does not change the form of the wave functions $\tilde{\psi}(\mathbf{r})$, i.e., $\tilde{\psi}_S(\mathbf{r}) = \tilde{\psi}_{PG}(\mathbf{r})$. It follows from equality (59) that for the solution with $E = 0$, the Hamiltonian for the PG metric coincides with the Hamiltonian for original Schwarzschild metric (39) up to a matrix functional factor:

$$\tilde{H}_{PG}(\mathbf{r}) = P_{PG}(r) \tilde{H}_S(\mathbf{r}). \tag{98}$$

The definition of Hamiltonian $\tilde{H}_{PG}(\mathbf{r})$ with tetrads (94) - (97) and with calculation of bispinor connectivities (5) shows that

$$P_{PG}(r) = \frac{1}{\sqrt{f_S}}\gamma^{\underline{0}} - \frac{\sqrt{r_0/r}}{\sqrt{f_S}}\gamma^{\underline{1}}.$$

By virtue of representation (26) (with the radial functions $F_{PG}(\rho)$ and $G_{PG}(\rho)$) and change (28), the components of the current density for tetrads (93) - (97) are

$$j^0 = \frac{1}{\sqrt{f_S}}\Big[\big(F_{PG}^*(\rho)F_{PG}(\rho) + G_{PG}^*(\rho)G_{PG}(\rho)\big)\xi^+(\theta)\xi(\theta) - i\sqrt{r_0/r} \times$$
$$\times \big(F_{PG}^*(\rho)G_{PG}(\rho) - F_{PG}(\rho)G_{PG}^*(\rho)\big)\xi^+(\theta)\xi(\theta)\Big],$$

$$j^\rho = -i\sqrt{f_S}\big(F_{PG}^*(\rho)G_{PG}(\rho) - F_{PG}(\rho)G_{PG}^*(\rho)\big)\xi^+(\theta)\xi(\theta),$$

$$j^\theta = -\frac{1}{\rho}\big(F_{PG}^*(\rho)G_{PG}(\rho) + F_{PG}(\rho)G_{PG}^*(\rho)\big)\xi^+(\theta)\sigma^2\xi(\theta),$$

$$j^\varphi = \frac{1}{\rho\sin\theta}\big(F_{PG}^*(\rho)G_{PG}(\rho) + F_{PG}(\rho)G_{PG}^*(\rho)\big)\xi^+(\theta)\sigma^1\xi(\theta).$$

Using Lorentz transformation (9), (10), (15), and (16), we transform tetrads (93) - (97) into tetrads (B.11) in the Schwinger gauge. The nonzero qualities $\Lambda^{\underline{\beta}}_\alpha(r)$ in formulas (9) and (10) are

$$\Lambda^{\underline{0}}_{\underline{0}} = \Lambda^{\underline{1}}_{\underline{1}} = \frac{1}{\sqrt{f_S}}; \quad \Lambda^{\underline{1}}_{\underline{0}} = \Lambda^{\underline{0}}_{\underline{1}} = -\frac{\sqrt{r_0/r}}{\sqrt{f_S}}. \tag{99}$$

Hamiltonian (98) after the transformation has form (92).

In the transformation process, the expression $\sqrt{f_S} = \sqrt{1 - r_0/r}$ appears in formulas (93) - (95), and (99); it is real in basic metric (39)(see condition (43)). It hence follows that the domain for the wave functions of Hamiltonian (92) is again the region $r \in (r_0, \infty)$. The Hilbert causality condition $g_{00} > 0$ for PG metric (91) also leads to the domain $r > r_0$.

Coordinate transformation (B.10) for tetrads (93) - (97) preserves the realness of the radial functions $F_{PG}(\rho)$ and $G_{PG}(\rho)$. Therefore, the radial and $\theta$ components of the current density vanish, and Hermiticity condition (19) for the Hamiltonian holds in the case of the PG metric.

According to (98), all the conclusion in Sec. 3 for original Schwarzschild metric (39) hold for the PG metric. In the PG gravitational field, there are degenerate stationary bound states of fermions with zero energy $E = 0$ and with the square-integrable spinor wave functions satisfying self-conjugate second-order equations.

The transition from Hamiltonian (98) to Hamiltonian (92) with tetrads in the Schwinger gauge is performed using a Lorentz transformation (see formulas (9), (10), (15), and (16)). In this case, the Dirac equations for the radial functions and the radial functions themselves become complex. The components of the current of the Dirac particles do not change under Lorentz transformations. Again, the radial and $\theta$ components of the current vanish (including the event horizon), and the Hermiticity condition according to expression (19) holds for the Hamiltonian.

After separation of variables in the Dirac equation with Hamiltonian (92), the system of equations for the complex radial functions $F_{PG-C}$ and $G_{PG-C}$ is

$$f_S \frac{dF_{PG-C}}{d\rho} + \left[\frac{1+\kappa}{\rho} - \frac{3}{4}\frac{2\alpha}{\rho^2} + i(1-\varepsilon)\sqrt{\frac{2\alpha}{\rho}}\right] F_{PG-C} - \left[1+\varepsilon+i\sqrt{\frac{2\alpha}{\rho}}\frac{1/4-\kappa}{\rho}\right] G_{PG-C} = 0,$$
$$f_S \frac{dG_{PG-C}}{d\rho} + \left[\frac{1-\kappa}{\rho} - \frac{3}{4}\frac{2\alpha}{\rho^2} - i(1+\varepsilon)\sqrt{\frac{2\alpha}{\rho}}\right] G_{PG-C} - \left[1-\varepsilon-i\sqrt{\frac{2\alpha}{\rho}}\frac{1/4+\kappa}{\rho}\right] F_{PG-C} = 0.$$
(100)

If the equalities $F_{PG-C}|_{\rho \to 2\alpha} = (\rho - 2\alpha)^s \sum_{k=0}^{\infty} f_k (\rho - 2\alpha)^k$ and $G_{PG-C}|_{\rho \to 2\alpha} = (\rho - 2\alpha)^s \sum_{k=0}^{\infty} g_k (\rho - 2\alpha)^k$ hold near the event horizon $(\rho \to 2\alpha)$, then the indicial equation for system (100) with $\varepsilon = 0$ leads to the solutions $s_1 = 0, s_2 = -1/2$. The solution $s_2 = -1/2$ corresponds to asymptotic formulas (52) for the radial real functions of the Dirac equation in the field of original Schwarzschild metric (39) and to the asymptotic formulas for radial functions of Hamiltonian (98) for the PG metric. Because Lorentz transformations do not lead to new physical consequences, the solution $s_1 = 0$ is mathematical artifact and is not used in what follows. We note that both solutions of the second-order equation lead to square-integrable wave functions with zero values at the event horizon.

To pass to the second-order equation, we write the expressions for $A_{PG}(\rho), B_{PG}(\rho), C_{PG}(\rho)$ and $D_{PG}(\rho)$:

$$A_{PG}(\rho) = -\frac{1}{f_S}\left(\frac{1+\kappa}{\rho} - \frac{3}{4}\frac{2\alpha}{\rho^2} + i(1-\varepsilon)\sqrt{\frac{2\alpha}{\rho}}\right), \quad B_{PG}(\rho) = \frac{1}{f_S}\left(1+\varepsilon+i\sqrt{\frac{2\alpha}{\rho}}\frac{\frac{1}{4}-\kappa}{\rho}\right),$$
$$C_{PG}(\rho) = \frac{1}{f_S}\left(1-\varepsilon-i\sqrt{\frac{2\alpha}{\rho}}\frac{\frac{1}{4}+\kappa}{\rho}\right), \quad D_{PG}(\rho) = -\frac{1}{f_S}\left(\frac{1-\kappa}{\rho} - \frac{3}{4}\frac{2\alpha}{\rho^2} - i(1+\varepsilon)\sqrt{\frac{2\alpha}{\rho}}\right).$$
(101)

According to formulas (32) - (37),

$$\Phi_{PG}(\rho) = g_{PG} F_{PG-C}(\rho), \; g_{PG}(\rho) = \exp\left[\frac{1}{2}\int A_F^{PG}(\rho') \, d\rho'\right], \quad (102)$$

$$A_F^{PG}(\rho) = -\frac{1}{B_{PG}}\frac{dB_{PG}}{d\rho} - A_{PG} - D_{PG}.$$

The second-order equation is

$$\frac{d^2\Phi_{PG}}{d\rho^2} + 2\left(E_{Schr} - U_{eff}^{PG}\right)\Phi_{PG} = 0, \quad (103)$$

$$E_{Schr} = \frac{1}{2}(\varepsilon^2 - 1), \quad (104)$$

$$U_{eff}^{PG} = -\frac{1}{4}\frac{1}{B_{PG}}\frac{d^2 B_{PG}}{d\rho^2} + \frac{3}{8}\left(\frac{1}{B_{PG}}\frac{dB_{PG}}{d\rho}\right)^2 - \frac{1}{4}(A_{PG} - D_{PG})\frac{1}{B_{PG}}\frac{dB_{PG}}{d\rho} + \frac{1}{4}\frac{d}{d\rho}(A_{PG} - D_{PG}) +$$
$$+ \frac{1}{8}(A_{PG} - D_{PG})^2 + \frac{1}{2}B_{PG} C_{PG} + E_{Schr}. \quad (105)$$

All the relationships (101) - (103) and (105) are complex. But the asymptotic relations considered above in formulas (54) - (56) with $\varepsilon = 0$ do not change and are localized on the real axis:

$$g_{PG}\big|_{\rho\to\infty} = \rho, \; g_{PG}\big|_{\rho\to 2\alpha} = (\rho - 2\alpha)^{3/4}, \; \Phi_{PG}\big|_{\rho\to\infty} = C_1 \varphi_1(\rho) \rho e^{-\rho\sqrt{1-\varepsilon^2}},$$
$$\Phi_{PG}\big|_{\rho\to 2\alpha} = L(\rho - 2\alpha)^{1/4}, \; U_{eff}^{PG}\big|_{\rho\to 2\alpha} = -\frac{3}{32}\frac{1}{(\rho - 2\alpha)^2}. \quad (106)$$

Equalities (106) imply that $\Phi_{PG}(\rho)$ is square integrable. The fermion wave function vanishes on the event horizon. Numerical calculations of the solution of second-order equation (103) confirm the existence of a degenerate stationary bound state with $\varepsilon = 0$. The calculated probability densities for detecting the fermion with $\varepsilon = 0$ and with complex eigenfunctions $\Phi_{PG}(\rho)$ for $\alpha \geq 1$ are close to the analogous probability densities calculated with the real functions $\Phi(\rho)$ of the second-order equation in the gravitational field of the original Schwarzschild metric. We will present the calculation results in another paper.

### 5. Lemaître-Finkelstein and Kruskal-Szekeres metrics

#### 5.1 Lemaître-Finkelstein solution

The square of the interval is given by the expression

$$ds^2 = dT^2 - \frac{dR^2}{[3(R-T)/2r_0]^{2/3}} - \left[\frac{3}{2}(R-T)\right]^{4/3} r_0^{2/3}\left(d\theta^2 + \sin^2\theta d\varphi^2\right). \quad (107)$$

According to formula (B.15), the domain of $T$ and $R$ is

$$R > T. \quad (108)$$

The quantities $-g$, $g_G$, and $\eta$ are defined by $-g = [3(R-T)/2]^2 r_0^2 \sin^2\theta$, $g_G = [3(R-T)/2]^2 r_0^2/R^4$, and

$$\eta_{LF} = (g_G)^{1/4} (g^{00})^{1/4} = \left( \left[\frac{3}{2}(R-T)\right]^2 \frac{r_0^2}{R^4} \right)^{1/4}. \tag{109}$$

The coordinate transformation, nonzero tetrads $\tilde{H}_{\underline{\alpha}}^\mu$, and matrices $\tilde{\gamma}^\alpha$ are given in Appendix B (see formulas (B.13 – B.17)).

The self-conjugate Hamiltonian in the $\eta$-representation with tetrads (B.16) is [25]

$$H_\eta = \gamma^{\underline{0}} m - i\gamma^{\underline{0}}\gamma^{\underline{1}} \left[\frac{3}{2r_0}(R-T)\right]^{1/3} \left(\frac{\partial}{\partial R} + \frac{1}{R}\right) - i\gamma^{\underline{0}}\gamma^{\underline{2}} \frac{1}{[3(R-T)/2]^{2/3} r_0^{1/3}} \times$$
$$\times \left(\frac{\partial}{\partial\theta} + \frac{1}{2}\mathrm{ctg}\,\theta\right) - i\gamma^{\underline{0}}\gamma^{\underline{3}} \frac{1}{[3(R-T)/2]^2 r_0^{1/3} \sin\theta} \frac{\partial}{\partial\varphi} - \frac{i}{2}\gamma^{\underline{0}}\gamma^{\underline{1}} \frac{\partial}{\partial R} \left[\frac{3}{2r_0}(R-T)\right]^{1/3}. \tag{110}$$

Under coordinate transformations (B.14), (B.15), the nonzero tetrads transformed according to (38) are

$$\left(H_{\underline{0}}^{\prime 0}\right)_{LF} = \frac{\partial T}{\partial t} \left(H_{\underline{0}}^0\right)_S = \frac{1}{\sqrt{f_S(R,T)}}, \tag{111}$$

$$\left(H_{\underline{0}}^{\prime 1}\right)_{LF} = \frac{\partial R}{\partial t} \left(H_{\underline{0}}^0\right)_S = \frac{1}{\sqrt{f_S(R,T)}}, \tag{112}$$

$$\left(H_{\underline{1}}^{\prime 0}\right)_{LF} = \frac{\partial T}{\partial r} \left(H_{\underline{1}}^1\right)_S = \frac{\sqrt{r_0/r(R,T)}}{\sqrt{f_S(R,T)}}, \tag{113}$$

$$\left(H_{\underline{1}}^{\prime 1}\right)_{LF} = \frac{\partial R}{\partial r} \left(H_{\underline{1}}^1\right)_S = \frac{1}{\sqrt{f_S(R,T)}\sqrt{r_0/r(R,T)}}, \tag{114}$$

$$\left(H_{\underline{2}}^{\prime 2}\right)_{LF} = \left(H_{\underline{2}}^2\right)_S = \frac{1}{r(R,T)}, \tag{115}$$

$$\left(H_{\underline{3}}^{\prime 3}\right)_{LF} = \left(H_{\underline{3}}^3\right)_S = \frac{1}{r(R,T)\sin\theta}. \tag{116}$$

In terms of the variables $R$ and $T$, according to equality (B.15),

$$f_S(R,T) = 1 - \frac{r_0}{r(R,T)} = 1 - \left(\frac{r_0}{3(R-T)/2}\right)^{2/3}. \tag{117}$$

Compared with tetrads (40), two additional nonzero tetrads (112) and (113) appear as a result of transformations (B.14).

Under coordinate transformations (B.14) and (B.15), the form of the wave functions for the Hamiltonian with tetrads (111) - (116) does not change. Representation (26) for the LF metric can be written as

$$\psi_{LF}(R,T) = \begin{pmatrix} F_{LF}(R,T)\xi(\theta) \\ -iG_{LF}(R,T)\sigma^3\xi(\theta) \end{pmatrix} e^{-iEt(R,T)} e^{im_\varphi \varphi}. \tag{118}$$

The function $t(R,T)$ can be determined from relations (B.14) and (B.15). By virtue of change (28) and representation (118), the components of the current density for tetrads (111) - (116) are

$$j^0 = \frac{1}{\eta_{LF}^2} \frac{1}{\sqrt{f_S(R,T)}} \left[ \left( F_{LF}^*(R,T)F_{LF}(R,T) + G_{LF}^*(R,T)G_{LF}(R,T) \right) \xi^+(\theta)\xi(\theta) - \right.$$
$$\left. -i\sqrt{r_0/r(R,T)} \left( F_{LF}^*(R,T)G_{LF}(R,T) - F_{LF}(R,T)G_{LF}^*(R,T) \right) \xi^+(\theta)\xi(\theta) \right], \tag{119}$$

$$j^\rho = \frac{1}{\eta_{LF}^2} \frac{1}{\sqrt{f_S(R,T)}} \left[ \left( F_{LF}^*(R,T)F_{LF}(R,T) + G_{LF}^*(R,T)G_{LF}(R,T) \right) \xi^+(\theta)\xi(\theta) - \right.$$
$$\left. -i\frac{1}{\sqrt{r_0/r(R,T)}} \left( F_{LF}^*(R,T)G_{LF}(R,T) - F_{LF}(R,T)G_{LF}^*(R,T) \right) \xi^+(\theta)\xi(\theta) \right], \tag{120}$$

$$j^\theta = -\frac{1}{\eta_{LF}^2} \frac{1}{r(R,T)} \left( F_{LF}^*(\rho)G_{LF}(\rho) + F_{LF}(\rho)G_{LF}^*(\rho) \right) \xi^+(\theta)\sigma^2\xi(\theta), \tag{121}$$

$$j^\varphi = -\frac{1}{\eta_{LF}^2} \frac{1}{r(R,T)\sin\theta} \left( F_{LF}^*(\rho)G_{LF}(\rho) + F_{LF}(\rho)G_{LF}^*(\rho) \right) \xi^+(\theta)\sigma^1\xi(\theta). \tag{122}$$

The quantity $\eta_{LF}$ in formelas (119) - (122) is defined by expression (109). The $\theta$ component of current density (121) vanishes in the whole domain of the wave functions. The second terms in the square brackets of expression (119) and (120) for $j^0$ and $j^\rho$ with the real radial functions $F_{LF}(\rho)$ and $G_{LF}(\rho)$ vanish. But the first terms in the square brackets of the expressions for $j^0$ and $j^\rho$ equal each other and are nonzero: $j^0 = j^\rho \neq 0$. Hermiticity condition (19) for the Hamiltonian is not satisfied. The Dirac Hamiltonians for the LF metric depend on time explicitly and are not Hermitian.

Using the Lorentz transformation (9), (10), (15), and (16), we now transform tetrads (111) - (116) to tetrads (B.16) in the Schwinger gauge. The nonzero quantities $\Lambda_{\underline{\alpha}}^{\underline{\beta}}(R,T)$ in expression (9) and (10) are given by

$$\Lambda_{\underline{0}}^{\underline{0}} = \Lambda_{\underline{1}}^{\underline{1}} = \frac{1}{\sqrt{f_S(R,T)}}, \quad \Lambda_{\underline{1}}^{\underline{0}} = \Lambda_{\underline{0}}^{\underline{1}} = -\sqrt{\frac{r_0}{r(R,T)}} \frac{1}{\sqrt{f_S(R,T)}}. \tag{123}$$

After the transformations, Hamiltonian (41) has the form (110).

The Hilbert causality condition $g_{00} > 0$ for LF metric (107) does not impose any restrictions on the domain for the wave functions of Hamiltonian (110). But in the transformation process, an expressions $\sqrt{f_S(R,T)} = \sqrt{1 - r_0/r(R,T)}$ appears in (111) - (114) and (123); it is real and positive in basic metric (39). It hence follows (see formula (117)) that in addition to condition (108), there is the restriction $R - T > 2r_0/3$ on the domain for the wave functions of Hamiltonian (110).

### 6.2 Kruskal-Szekeres solution

The square of the interval is given by

$$ds^2 = f^2 dv^2 - f^2 du^2 - (r(u,v))^2 (d\theta^2 + \sin^2 d\varphi^2), \tag{124}$$

where $(f(u,v))^2 = (4r_0^3/r(u,v)) e^{-r(u,v)/r_0}$ is a function of $u^2 - v^2$.

The quantities $(-g)$, $g_G$ and $\eta_{KS}$ are $-g = (f(u,v))^4 (r(u,v))^4 \sin^2 \theta$, $g_G = (f(u,v))^4 (r(u,v))^4 / u^4$ and

$$\eta_{KS} = (g_G \cdot g^{00})^{1/4} = (f(u,v))^{1/2} \frac{r(u,v)}{u}. \tag{125}$$

The coordinate transformations, the nonzero tetrads $\tilde{H}_{\underline{\alpha}}^{\mu}$, and matrices $\tilde{\gamma}^{\alpha}$ are given in Appendix B (see formulas (B.18) – (B.23)).

The self-conjugate Hamiltonian in the $\eta$ - representation with tetrads (B.22) is

$$H_\eta = \gamma^{\underline{0}} f(u,v) m - i \gamma^{\underline{0}} \gamma^{\underline{1}} \left( \frac{\partial}{\partial u} + \frac{1}{u} \right) - i \gamma^{\underline{0}} \gamma^{\underline{2}} \frac{f(u,v)}{r(u,v)} \left( \frac{\partial}{\partial \theta} + \frac{1}{2} \operatorname{ctg} \theta \right) -$$
$$- i \gamma^{\underline{0}} \gamma^{\underline{3}} \frac{f(u,v)}{r(u,v) \sin \theta} \frac{\partial}{\partial \varphi}. \tag{126}$$

The nonzero tetrads transformed under coordinate transformations (B.19) - (B.21) according to formulas (38) are given by

$$\left( H_{\underline{0}}'^0 \right)_{KS} = \frac{\partial v}{\partial t} \left( H_{\underline{0}}^0 \right)_S = \operatorname{ch}\left( \frac{t(u,v)}{2 r_0} \right) \frac{1}{2 r_0} \sqrt{\frac{r(u,v)}{r_0}} \exp \frac{r(u,v)}{2 r_0}, \tag{127}$$

$$\left( H_{\underline{0}}'^1 \right)_{KS} = \frac{\partial u}{\partial t} \left( H_{\underline{0}}^0 \right)_S = \operatorname{sh}\left( \frac{t(u,v)}{2 r_0} \right) \frac{1}{2 r_0} \sqrt{\frac{r(u,v)}{r_0}} \exp \frac{r(u,v)}{2 r_0}, \tag{128}$$

$$\left( H_{\underline{1}}'^0 \right)_{KS} = \frac{\partial v}{\partial r} \left( H_{\underline{1}}^1 \right)_S = \operatorname{sh}\left( \frac{t(u,v)}{2 r_0} \right) \frac{1}{2 r_0} \sqrt{\frac{r(u,v)}{r_0}} \exp \frac{r(u,v)}{2 r_0}, \tag{129}$$

$$\left(H_{\underline{1}}^{\prime 1}\right)_{KS} = \frac{\partial u}{\partial r}\left(H_{\underline{1}}^{1}\right)_{S} = \operatorname{ch}\left(\frac{t(u,v)}{2r_0}\right)\frac{1}{2r_0}\sqrt{\frac{r(u,v)}{r_0}}\exp\frac{r(u,v)}{2r_0} \tag{130}$$

$$\left(H_{\underline{2}}^{\prime 2}\right)_{KS} = \left(H_{\underline{2}}^{2}\right)_{S} = \frac{1}{r(u,v)}, \quad \left(H_{\underline{3}}^{\prime 3}\right)_{KS} = \left(H_{\underline{3}}^{3}\right)_{S} = \frac{1}{r(u,v)\sin\theta}. \tag{131}$$

Compared with tetrads (40), two additional nonzero tetrads (128) and (129) appear as result of transformations (B.19) – (B.21).

The form of the wave functions for the Hamiltonian with tetrads (127) - (131) does not change under coordinate transformations (B.19), (B.20). Representation (26) for the KSz metric can be written as

$$\psi_{KS}(u,v) = \begin{pmatrix} F_{KS}(u,v)\xi(\theta) \\ -iG_{KS}(u,v)\sigma^3\xi(\theta) \end{pmatrix} e^{-iEt(u,v)} e^{im_\varphi\varphi}. \tag{132}$$

The function $t(u,v)$ is defined by the equality (B.20). By virtue of (28) and representation (132), the components of the current density for tetrads (127) - (131) are

$$j^0 = \frac{1}{\eta_{KS}^2}\frac{1}{2r_0}\sqrt{\frac{r(u,v)}{r_0}}\exp\frac{r(u,v)}{2r_0}\left[\operatorname{ch}\left(\frac{t(u,v)}{2r_0}\right)\left(F_{KS}^*(u,v)F_{KS}(u,v) + G_{KS}^*(u,v)G_{KS}(u,v)\right)\times\right.$$
$$\left.\times\xi^+(\theta)\xi(\theta) - i\operatorname{sh}\left(\frac{t(u,v)}{2r_0}\right)\left(F_{KS}^*(u,v)G_{KS}(u,v) - F_{KS}(u,v)G_{KS}^*(u,v)\right)\xi^+(\theta)\xi(\theta)\right], \tag{133}$$

$$j^\rho = \frac{1}{\eta_{KS}^2}\frac{1}{2r_0}\sqrt{\frac{r(u,v)}{r_0}}\exp\frac{r(u,v)}{2r_0}\left[\operatorname{sh}\left(\frac{t(u,v)}{2r_0}\right)\left(F_{KS}^*(u,v)F_{KS}(u,v) + G_{KS}^*(u,v)G_{KS}(u,v)\right)\times\right.$$
$$\left.\times\xi^+(\theta)\xi(\theta) - i\operatorname{ch}\left(\frac{t(u,v)}{2r_0}\right)\left(F_{KS}^*(u,v)G_{KS}(u,v) - F_{KS}(u,v)G_{KS}^*(u,v)\right)\xi^+(\theta)\xi(\theta)\right], \tag{134}$$

$$j^\theta = \frac{1}{\eta_{KS}^2}\frac{1}{r(u,v)}\left(F_{KS}^*(u,v)G_{KS}(u,v) + F_{KS}(u,v)G_{KS}^*(u,v)\right)\xi^+(\theta)\sigma^2\xi(\theta), \tag{135}$$

$$j^\varphi = \frac{1}{\eta_{KS}^2}\frac{1}{r(u,v)\sin\theta}\left(F_{KS}^*(u,v)G_{KS}(u,v) + F_{KS}(u,v)G_{KS}^*(u,v)\right)\xi^+(\theta)\sigma^1\xi(\theta). \tag{136}$$

The quantity $\eta_{KS}$ in formulas (133) - (136) is defined by expression (125). The $\theta$ component of current density (135) vanishes. The second terms in the square brackets of expressions (133) and (134) for the components $j^0$ and $j^\rho$ with the real radial functions $F_{KS}(\rho)$ and $G_{KS}(\rho)$ vanish. But the first term in the square brackets of the expression for $j^\rho$ is nonzero. The Dirac Hamiltonians for the KSz metric explicitly and are not Hermitian.

We transforme tetrads (127) - (131) into tetrads (B.22) in the Schwinger gauge using Lorentz transformations (9), (10), (15) and (16). The nonzero quantities $\Lambda_\alpha^\beta(u,v)$ in (9), (10) are defined as

$$\Lambda_{\underline{0}}^0 = \Lambda_{\underline{1}}^1 = \frac{1}{f}\text{ch}\left(\frac{t(u,v)}{2r_0}\right)\frac{1}{2r_0}\sqrt{\frac{r(u,v)}{r_0}}\exp\frac{r(u,v)}{2r_0};$$

$$\Lambda_{\underline{0}}^1 = \Lambda_{\underline{1}}^0 = -\frac{1}{f}\text{sh}\left(\frac{t(u,v)}{2r_0}\right)\frac{1}{2r_0}\sqrt{\frac{r(u,v)}{r_0}}\exp\frac{r(u,v)}{2r_0}.$$

Basic Hamiltonian (41) after the transformation has form (126).

The Hilbert causality condition $g_{00} > 0$ for KSz metric (124) does not restrict the domain for the wave functions of Hamiltonian (126). There are also no restrictions on obtaining tetrads (127) - (131) in the first step of transformation and on performing the Lorentz transformation in the second step of transformation of basic Hamiltonian (41). But a restriction of the domain for the wave functions of the transformed Hamiltonian arises when we introduce the new variables $(u,v)$. There is an expression $f_S = 1 - r_0/r(u,v)$ in equalities (B.19) and (B.20) that is real and positive in basic metric (39). It hence follows (see formulas (B.19) and (B.20)) that the conditions $u^2 > v^2$ and $u^2 \neq v^2 \neq 0$ must hold for the domain in the coordinates $(u,v)$. The right quadrant $u > |v|$ and the left quadrant $u < -|v|$ on the plane $(u,v)$ represent the domain for the wave functions of Hamiltonian (126). The lines $u = \pm v$ and the point $u = v = 0$ do not belong to the sought domain.

In contrast to the basic Hamiltonian (41), KSz metric (124) is nonstationary, and Hamiltonian (126) in terms of $(u,v)$ depends explicitly on the time coordinate $v$.

The analysis of the nonstationary LF and KSz metrics leads to the conclusion that these metrics are not equivalent to the static and starionary metrics. Nevertheless, all observable physical effects must be the same for any choice of the coordinate. The equivalence of the metrics can be restored if we use and evolution operator in the direction of the timelike Killing vector (see Appendix C). In that case, all our obtained conclusions for the static and stationary metric can also be reproduced for the nonstationary LF and KSz metrics.

### 7. Conclusion

It is known that quasistationary bound states of fermions decaying in time can exist when complex radial wave functions are used in the Dirac equation in the Schwarzschild field. These states are characterized by complex energies.

When real radial functions are used in the Dirac equation, the situation changes qualitatively [20]. In this case, the radial currents of Dirac particles vanish over the whole domain of the wave functions. The domain for the wave functions is $r > r_0$. But the radial functions are square nonintegrable as $r \to r_0$, and the regime of a particle "fall" on the event horizon is realized for nonzero values of the particle energy. The square integrability of the redial functions near the event horizon can be restored if the fermion motion is described by self-conjugate second-order equations with spinor wave functions [20]. Solving the second-order equation, we can obtain a degenerate stationary bound state of the massive fermion with zero energy. Normalized eigenfunctions vanishing on the event horizon correspond to this state.

We have analyzed the quantum mechanical equivalence analysis of the metrics of centrally symmetric uncharged gravitational fields with respect to fermions. We considered the static Schwarzschild metrics in spherical [1] and isotropic [2] coordinates, the stationary EF [3], [4] and PG metrics [5], [6], and the nonstationary LF [4], [7] and KSz metrics [8], [9]. All these metrics were obtained from the solution [1] by appropriate coordinate transformations.

For all the metrics, we obtain self-conjugate Hamiltonians with the plane scalar product of the wave functions and with tetrad vectors in the Schwinger gauge. In addition, the same Hamiltonians were obtained using direct two-step transformations of basic Hamiltonian (41) for the Schwarzschild field in the coordinates $(t, r, \theta, \varphi)$. First, according to the coordinate transformations for the considered metrics, we transformed basic Hamiltonian (41) with tetrads (40). If necessary, we then applied Lorentz transformations (9), (10), (15), and (16) to pass to tetrads in the Schwinger gauge.

For the considered metrics and Hamiltonians, we analyzed the domains for the wave functions of the Dirac equation, the Hermiticity of the Hamiltonians $\left((\Phi, H\Psi) = (H\Phi, \Psi)\right)$, and the possibility of the existence of a solution of the second-order equation corresponding to a degenerate stationary bound state of half-spin particles with a zero energy. As a result of this analysis, we obtained several conclusions.

1. For the basic Schwarzschild metric in the spherical coordinates $(t, r, \theta, \varphi)$ with real radial functions in the Dirac equation, the domain of wave functions is restricted by the condition

$$r > r_0. \tag{137}$$

For all other considered metrics, condition (137) in new variables are as follows:

- In the static Schwarzschild metric in isotropic coordinates, we have

$$R > \frac{r_0}{4}. \tag{138}$$

- In the stationary EF and PG metrics, we have

$$r > r_0. \qquad (139)$$

- In the nonstationary LF metric, we have

$$R - T > \frac{2}{3} r_0. \qquad (140)$$

- In the nonstationary KSz metric, we have

$$u > |v| > 0, \; u < -|v| < 0. \qquad (141)$$

From inequalities (138) - (141), we see that the "event horizon" $r_0$ in basic Schwarzschild metric (39) appears in the new coordinates in all considered metrics. The domains of wave functions for all metrics obtained by the coordinate transformations with condition (137) for the basic metric (39) do not include the space region under the event horizon nor the singularity at the coordinate origin. All these metrics in domains (137) - (141) are mutually related by analytical coordinate transformations. In these domains, we can introduce the background flat space whose metric $(g_c)_{\alpha\beta}$ in the appropriate coordinates is necessary for pseudo-Hermitian formulation of the solution to the Dirac equation.

2. Hamiltonian (61) for the static Schwarzschild metric in isotropic coordinates is Hermitian. Similarly, Hamiltonians (82), (88), (92), and (98) for the stationary EF and PG metrics are also Hermitian. After the passage from the Dirac equation to the self-conjugate second-order equation with effective potentials, there is a degenerate stationary bound state of zero-energy fermions for all studied metrics. The eigenfunctions corresponding to the eigenvalue $E = 0$ vanishes at the event horizon. The probability densities for detecting half-spin particles with fixed $j$ and $l$ in the dimensionless coordinates are the same for all static and stationary metrics.

Using the evolution operator along the timelike Killing vector, we can reproduce all the conclusions in Sec. 2 for the nonstationary LF and KSz merics.

Therefore, if real radial functions of the Dirac equation and of the second-order equation for the basic Schwarzschild metric with the coordinates $(t, r, \theta, \varphi)$ are used, all considered static, stationary, and nonstationary metrics are equivalent. The singularities of the effective potentials of the second-order equation near the event horizons are preserved for all considered metrics. There are regular stationary solutions of the second-order equation with zero-energy fermions for all metrics.

The atomic system with stationary bound states of the half-spin particles can be candidates for the role of dark matter particles [20].

**Appendix A: Effective potential of the Schwarzschild field in the Schrodinger-type equation**

According to formulas (36), (37), and (53) we can obtain the equalities

$$\frac{3}{8}\frac{1}{B^2}\left(\frac{dB}{d\rho}\right)^2 = \frac{3}{8}\left(\frac{2\alpha}{\rho(\rho-2\alpha)} - \frac{\alpha}{\varepsilon\rho(\rho(\rho-2\alpha))^{1/2} + \rho(\rho-2\alpha)}\right)^2, \qquad (A.1)$$

$$-\frac{1}{4}\frac{1}{B}\frac{d^2 B}{d\rho^2} = -\frac{2\alpha^2}{\rho^2(\rho-2\alpha)^2} - \frac{\alpha}{\rho^2(\rho-2\alpha)} + \frac{5}{4}\frac{\alpha^2}{\varepsilon\rho^{5/2}(\rho-2\alpha)^{3/2} + \rho^2(\rho-2\alpha)^2} +$$

$$+\frac{\alpha}{2\left[\varepsilon\rho^{5/2}(\rho-2\alpha)^{1/2} + \rho^2(\rho-2\alpha)\right]},$$

$$\frac{1}{4}\frac{d}{d\rho}(A-D) = \frac{\kappa(\rho-\alpha)}{2\rho^{3/2}(\rho-2\alpha)^{3/2}},$$

$$-\frac{1}{4}\frac{(A-D)}{B}\frac{dB}{d\rho} = -\frac{\alpha\kappa}{\rho^{3/2}(\rho-2\alpha)^{3/2}} + \frac{\alpha\kappa}{2\left[\varepsilon\rho^2(\rho-2\alpha) + \rho^{3/2}(\rho-2\alpha)^{3/2}\right]},$$

$$\frac{1}{8}(A-D)^2 = \frac{\kappa^2}{2\rho(\rho-2\alpha)}, \quad \frac{1}{2}BC = -\frac{1}{2}\frac{\rho^2\varepsilon^2}{(\rho-2\alpha)^2} + \frac{1}{2}\frac{\rho}{\rho-2\alpha}. \qquad (A5)$$

The sum of the expressions for $E_{Schr}$ and (A.1) leads to the sought expression for the effective potential $U_{eff}^F$. The asymptotics form is $U_{eff}^F(\varepsilon=0)\big|_{\rho\to 2\alpha} = -3\big/\left(32(\rho-2\alpha)^2\right)$.

**Appendix B: Coordinate transformations, nonzero components of tetrad vectors $\tilde{H}_{\underline{\alpha}}^{\mu}$ in the Schwinger gauge and Dirac matrices $\left(\tilde{\gamma}^{\alpha} = \tilde{H}_{\underline{\beta}}^{\alpha}\gamma^{\underline{\beta}}\right)$ with world indices**

1. *Schwarzschild metric in isotropic coordinates:* We use the coordinates

$$(t, R, \theta, \varphi). \qquad (B.1)$$

The coordinate transformation is

$$r = R\left(1 + \frac{r_0}{4R}\right)^2, \quad R = \frac{1}{2}\left[\left(r - \frac{r_0}{2}\right) \pm \sqrt{r(r-r_0)}\right], \quad dR = \frac{dr}{\left(1 - r_0^2/16R^2\right)}. \qquad (B.2)$$

An ambiguilty in the definition of the new coordinate $R$ follows from equalities (B.2).

The nonzero components of the tetrad vectors $\tilde{H}_{\underline{\alpha}}^{\mu}$ in the Schwinger gauge are defined as

$$\tilde{H}^0_{\underline{0}} = \frac{1+r_0/4R}{1-r_0/4R}, \qquad \tilde{H}^1_{\underline{1}} = \frac{1}{(1+r_0/4R)^2},$$
$$\tilde{H}^2_{\underline{2}} = \frac{1}{(1+r_0/4R)^2}\frac{1}{R}, \quad \tilde{H}^3_{\underline{3}} = \frac{1}{(1+r_0/4R)^2}\frac{1}{R\sin\theta}. \tag{B.3}$$

The matrices $\tilde{\gamma}^\alpha$ are

$$\tilde{\gamma}^0 = \frac{1+r_0/4R}{1-r_0/4R}\gamma^{\underline{0}}, \qquad \tilde{\gamma}^1 = \frac{1}{(1+r_0/4R)^2}\gamma^{\underline{1}},$$
$$\tilde{\gamma}^2 = \frac{1}{R(1+r_0/4R)^2}\gamma^{\underline{2}}, \quad \tilde{\gamma}^3 = \frac{1}{R\sin\theta(1+r_0/4R)^2}\gamma^{\underline{3}}. \tag{B.4}$$

2. *Eddington-Finkelstein metric.* Here, we choose the coordinates

$$(T,r,\theta,\varphi). \tag{B.5}$$

The coordinate transformation is

$$dT = dt + \frac{r_0}{r}\frac{dr}{f_S}. \tag{B.6}$$

The nonzero tetrads $\tilde{H}^\mu_{\underline{\alpha}}$ in the Schwinger gauge are defined as

$$\tilde{H}^0_{\underline{0}} = \sqrt{1+\frac{r_0}{r}}, \quad \tilde{H}^1_{\underline{0}} = -\frac{r_0/r}{\sqrt{1+r_0/r}}, \quad \tilde{H}^1_{\underline{1}} = \frac{1}{\sqrt{1+r_0/r}},$$
$$\tilde{H}^2_{\underline{2}} = r^{-1}, \quad \tilde{H}^3_{\underline{3}} = (r\sin\theta)^{-1}. \tag{B.7}$$

The matrices $\tilde{\gamma}^\alpha$ are

$$\tilde{\gamma}^0 = \sqrt{1+\frac{r_0}{r}}\gamma^{\underline{0}}, \quad \tilde{\gamma}^1 = -\frac{r_0/r}{\sqrt{1+r_0/r}}\gamma^{\underline{0}} + \frac{1}{\sqrt{1+r_0/r}}\gamma^{\underline{1}}, \quad \tilde{\gamma}^2 = \frac{\gamma^{\underline{2}}}{r}, \quad \tilde{\gamma}^3 = \frac{\gamma^{\underline{3}}}{r\sin\theta}. \tag{B.8}$$

3. *The Painlevé-Gullstrand metric:* We choose the oordinates

$$(T,r,\theta,\varphi). \tag{B.9}$$

The coordinate transformation is defined by

$$dT = dt + \sqrt{r_0/r}(f_S)^{-1}dr. \tag{B.10}$$

The nonzero tetrads $\tilde{H}^\mu_{\underline{\alpha}}$ in the Schwinger gauge are

$$\tilde{H}^0_{\underline{0}} = 1, \quad \tilde{H}^1_{\underline{0}} = -\sqrt{r_0/r}, \quad \tilde{H}^1_{\underline{1}} = 1, \quad \tilde{H}^2_{\underline{2}} = r^{-1}, \quad \tilde{H}^3_{\underline{3}} = (r\sin\theta)^{-1}. \tag{B.11}$$

The matrices $\tilde{\gamma}^\alpha$ are

$$\tilde{\gamma}^0 = \gamma^{\underline{0}}, \quad \tilde{\gamma}^1 = -\sqrt{\frac{r_0}{r}}\gamma^{\underline{0}} + \gamma^{\underline{1}}, \quad \tilde{\gamma}^2 = \frac{1}{r}\gamma^{\underline{2}}, \quad \tilde{\gamma}^3 = \frac{1}{r\sin\theta}\gamma^{\underline{3}}. \tag{B.12}$$

4. *The Lemaître-Finkelstein metric:* We choose the coordinates

$$(T, R, \theta, \varphi). \tag{B.13}$$

The coordinate transformations are

$$dT = dt + \frac{\sqrt{r_0/r}}{f_S}dr, \quad dR = dt + \frac{dr}{f_S\sqrt{r_0/r}}, \tag{B.14}$$

$$R = T + \frac{2}{3}\frac{r^{3/2}}{r_0^{1/2}}, \quad r = \left[\frac{3}{2}(R-T)\right]^{2/3} r_0^{1/3}. \tag{B.15}$$

The nonzero components of the tetrad vectors $\tilde{H}^\mu_{\underline{\alpha}}$ in the Schwinger gauge are

$$\tilde{H}^0_{\underline{0}} = 1, \quad \tilde{H}^1_{\underline{1}} = \left[\frac{3}{2r_0}(R-T)\right]^3, \quad \tilde{H}^2_{\underline{2}} = \frac{1}{\left[\frac{3}{2}(R-T)\right]^{2/3} r_0^{1/3}}, \quad \tilde{H}^3_{\underline{3}} = \frac{1}{\left[\frac{3}{2}(R-T)\right]^{2/3} r_0^{1/3} \sin\theta}. \tag{B.16}$$

The matrices $\tilde{\gamma}^\alpha$ are

$$\tilde{\gamma}^0 = \gamma^{\underline{0}}, \quad \tilde{\gamma}^1 = \left[\frac{3}{2r_0}(R-T)\right]^{1/3} \gamma^{\underline{1}}, \quad \tilde{\gamma}^2 = \frac{1}{\left[\frac{3}{2}(R-T)\right]^{2/3} r_0^{1/3}} \gamma^{\underline{2}},$$

$$\tilde{\gamma}^3 = \frac{1}{\left[\frac{3}{2}(R-T)\right]^{2/3} r_0^{1/3} \sin\theta} \gamma^{\underline{3}}. \tag{B.17}$$

5. *The Kruskal-Szekeres metric:* We choose the coordinates

$$(v, u, \theta, \varphi). \tag{B.18}$$

The coordinate transformations are

$$u = \sqrt{\frac{r}{r_0}}\sqrt{f_S}\, e^{r/2r_0} \operatorname{ch}\frac{t}{2r_0}, \quad v = \sqrt{\frac{r}{r_0}}\sqrt{f_S}\, e^{r/2r_0} \operatorname{sh}\frac{t}{2r_0}, \tag{B.19}$$

$$\frac{r}{r_0}\sqrt{f_S}\, e^{r/2r_0} = u^2 - v^2, \quad \frac{t}{2r_0}\operatorname{arcth}\frac{v}{u} = \frac{1}{2}\operatorname{arcth}\frac{2uv}{u^2+v^2}, \tag{B.20}$$

$$du = \frac{1}{2r_0}\sqrt{r/r_0}\sqrt{f_S}\, e^{r/2r_0}\operatorname{sh}\frac{t}{2r_0}dt + \frac{1}{2r_0}\frac{\sqrt{r/r_0}}{\sqrt{f_S}}\exp\frac{r}{2r_0}\operatorname{ch}\frac{t}{2r_0}dr,$$

$$dv = \frac{1}{2r_0}\sqrt{r/r_0}\sqrt{f_S}\, e^{r/2r_0}\operatorname{ch}\frac{t}{2r_0}dt + \frac{1}{2r_0}\frac{\sqrt{r/r_0}}{\sqrt{f_S}}e^{r/2r_0}\operatorname{sh}\frac{t}{2r_0}dr. \tag{B.21}$$

The nonzero components of the tetrad vectors $\tilde{H}^\mu_{\underline{\alpha}}$ in the Schwinger gauge are

$$\tilde{H}_{\underline{0}}^{0} = \frac{1}{f(u,v)}, \quad \tilde{H}_{\underline{1}}^{1} = \frac{1}{f(u,v)}, \quad \tilde{H}_{\underline{2}}^{2} = \frac{1}{r(u,v)}, \quad \tilde{H}_{\underline{3}}^{3} = \frac{1}{r(u,v)\sin\theta}. \tag{B.22}$$

The matrices $\tilde{\gamma}^{\alpha}$ are

$$\tilde{\gamma}^{0} = \frac{\gamma^{0}}{f}, \quad \tilde{\gamma}^{1} = \frac{\gamma^{1}}{f}, \quad \tilde{\gamma}^{2} = \frac{\gamma^{2}}{r(u,v)}, \quad \tilde{\gamma}^{3} = \frac{\gamma^{3}}{r(u,v)\sin\theta}. \tag{B.23}$$

**Appendix C: Evolution equations for nonstationary Lemaître-Finkelstein and Kruskal-Szekeres metrics**

1. *Lemaître-Finkelstein metric:* The coordinate transformations are

$$(t, r, \theta, \varphi) \to (T_{LF}, R_{LF}, \theta, \varphi), \tag{C.1}$$

$$dT_{LF} = dt + \frac{\sqrt{r_0/r}}{f_S} dr, \quad T_{LF} = t + \varphi_{LF}(r) = t + \int \frac{\sqrt{r_0/r}}{f_S} dr, \tag{C.2}$$

$$dR_{LF} = dt + \frac{dr}{\sqrt{r_0/r}\, f_S}, \quad R_{LF} = t + \chi_{LF}(r) = t + \int \frac{dr}{\sqrt{r_0/r}\, f_S}. \tag{C.3}$$

Under coordinate transformations (C.1) - (C.3), the wave functions of the base Schwarzschild metric remain unchanged except the corresponding change of variables. By virtue of formula (C.2),

$$\psi_{LF} = \tilde{\psi}_S(r,\theta,\varphi) e^{-iEt(T_{LF}, R_{LF})} = \tilde{\psi}_S(r,\theta,\varphi) e^{-iE(T_{LF} - \varphi_{LF}(r))}. \tag{C.4}$$

By virtue of (C.3)

$$\psi_{LF} = \tilde{\psi}_S(r,\theta,\varphi) e^{-iEt(T_{LF}, R_{LF})} = \tilde{\psi}_S(r,\theta,\varphi) e^{-iE(T_{LF} - \chi_{LF}(r))}. \tag{C.5}$$

It hence follows that

$$T_{LF} - \varphi_{LF}(r) = R_{LF} - \chi_{LF}(r), \quad R_{LF} - T_{LF} = \int \frac{1}{f_S} \left(\frac{1}{\sqrt{r_0/r}} - \sqrt{\frac{r_0}{r}}\right) dr = \frac{2}{3} \frac{r^{3/2}}{r_0^{1/2}},$$

$$r = r_0^{1/3} \left[\frac{3}{2}(R_{LF} - T_{LF})\right]^{2/3}. \tag{C.6}$$

In expressions (C.4) and (C.5) and below, we assume the dependence $r(T_{LF}, R_{LF})$ established in formula (C.6).

For the LF metric, the timelike Killing vector $\eta_{LF}^{\alpha} = (1,1,0,0)$. The evolution equation in the direction of the Killing vector is

$$i\left(\frac{\partial r}{\partial T_{LF}} + \frac{\partial r}{\partial R_{LF}}\right) \psi_{LF} = 2\tilde{H}_S \psi_{LF}. \tag{C.7}$$

By virtue of relations (C.2) - (C.6), we can write Eq. (C.7) as

$$\left[2E\tilde{\psi}_S(r,\theta,\varphi)+i\frac{\partial\tilde{\psi}_S}{\partial r}\left(\frac{\partial r}{\partial T_{LF}}+\frac{\partial r}{\partial R_{LF}}\right)\right]e^{-iEt(T_{LF},R_{LF})}=2\tilde{H}_S\tilde{\psi}_S(r,\theta,\varphi)e^{-iEt(T_{LF},R_{LF})}.$$

It follows from (C.6) that $\partial r/\partial T_{LF}=-\partial r/\partial R_{LF}$, and we finally obtain

$$E\tilde{\psi}_S(r,\theta,\varphi)=e^{iEt(T_{LF},R_{LF})}\tilde{H}_S e^{-iEt(T_{LF},R_{LF})}.$$

Taking the form of base Hamiltonian (41) into account, we obtain

$$E\tilde{\psi}_S(r,\theta,\varphi)=\left(\tilde{H}_S-Ef_S\left(\frac{\partial t}{\partial T_{LF}}\frac{\partial T_{LF}}{\partial r}+\frac{\partial t}{\partial R_{LF}}\frac{\partial R_{LF}}{\partial r}\right)\gamma^0\gamma^1\right)\tilde{\psi}_S(r,\theta,\varphi),$$

$$E\left(1+\frac{1+r_0/r}{\sqrt{r_0/r}}\gamma^0\gamma^1\right)\tilde{\psi}_S(r,\theta,\varphi)=\tilde{H}_S\tilde{\psi}_S(r,\theta,\varphi).$$
(C.8)

After separation of the variables, the term proportional to $E(1+r_0/r)/\sqrt{r_0/r}$ turns out to be the only imaginary term in each of the equations for the real radial functions. As a result, Eq. (C.8) has the unique solution $E=0$ and $\tilde{H}_S\tilde{\psi}_S(r,\theta,\varphi)=0$. All the conclusions for the basic Schwarzschild metric also hold for the LF metric. There is a strong connection between the Schwarzschild coordinate $r$ and the difference $R_{LF}-T_{LF}$ (see formula (C.6)).

2. *Kruskal-Szekeres metric:* The coordinate transformations are

$$(t,r,\theta,\varphi)\to(v,u,\theta,\varphi).$$
(C.9)

The explicit forms of the transformations are given in formulas (B.19) - (B.21). Under coordinate transformations (C.9), the wave functions remain unchanged except the corresponding change of variables:

$$\psi_{KS}=\tilde{\psi}_S(r(v,u),\theta,\varphi)e^{-iEt(v,u)}.$$
(C.10)

The dependence $r(v,u)$ and $t(v,u)$ are defined by equalities (B.20). The timelike vector for the KSz metric equals $\eta^\alpha=(\eta^0,\eta^1,0,0)$, where

$$\eta^0=\frac{\partial v}{\partial t}=\frac{1}{2r_0}\sqrt{\frac{r}{r_0}}\sqrt{f_S}\,e^{r/2r_0}\operatorname{ch}\frac{t}{2r_0},\quad \eta^1=\frac{\partial u}{\partial t}=\frac{1}{2r_0}\sqrt{\frac{r}{r_0}}\sqrt{f_S}\,e^{r/2r_0}\operatorname{sh}\frac{t}{2r_0}.$$
(C.11)

The evolution equation in the Killing vector direction is

$$i\left(\eta^0\frac{\partial}{\partial v}+\eta^1\frac{\partial}{\partial u}\right)\psi_{KS}=2\tilde{H}_S\psi_{KS}.$$

By virtue of formulas (C.10) and (C.11), we obtain

$$\left[E\tilde{\psi}_S(r(v,u),\theta,\varphi)+i\frac{\partial\tilde{\psi}_S}{\partial r}\left(\eta^0\frac{\partial r}{\partial v}+\eta^1\frac{\partial r}{\partial u}\right)\right]e^{-iEt(v,u)}=\tilde{H}_S\tilde{\psi}_S(r(v,u),\theta,\varphi)e^{-iEt(v,u)}.$$

By virtue of equality (B.20) $\eta^0\,\partial r/\partial v=-\eta^1\,\partial r/\partial u$, and

$$E\tilde{\psi}_S\left(r(v,u),\theta,\varphi\right) = e^{iEt(v,u)}\tilde{H}_S e^{-iEt(v,u)}\tilde{\psi}_S\left(r(v,u),\theta,\varphi\right), \quad \text{(C.12)}$$

$$E\left[1+\gamma^0\gamma^1 f_S\left(\frac{\partial t(v,u)}{\partial v}\frac{\partial v}{\partial r}+\frac{\partial t(v,u)}{\partial u}\frac{\partial u}{\partial r}\right)\right]\tilde{\psi}_S\left(r(v,u),\theta,\varphi\right) = \tilde{H}_S\tilde{\psi}_S\left(r(v,u),\theta,\varphi\right). \quad \text{(C.13)}$$

The expression

$$\left(\frac{\partial t}{\partial v}\frac{\partial v}{\partial r}+\frac{\partial t}{\partial u}\frac{\partial u}{\partial r}\right)$$

is defined according to formula (B.19).

After separation of the variables, the term in (C.13) becomes proportional to

$$E\gamma^0\gamma^1 f_S\left(\frac{\partial t}{\partial v}\frac{\partial v}{\partial r}+\frac{\partial t}{\partial u}\frac{\partial u}{\partial r}\right)$$

and turns out to be the only imaginary term in each of the two equations for the real radial functions. As a result, Eq. (C.13) has the unique solution $E=0$ and $\tilde{H}_S\tilde{\psi}_S\left(r(v,u),\theta,\varphi\right)=0$. All the conclusions for the basic Schwarzschild metric also hold for the nonstationary KSz metric. The relations between the Schwarzschild coordinates $r$ and $t$ and the KSz coordinates $v$ and $u$ are defined by formula (B.20).

**Acknowledgements**

The authors thank A.L. Novoselova for the essential technical assistance in preparing this paper.